\documentclass[sigconf]{acmart}

\usepackage{booktabs} 
\usepackage{graphicx}  
\usepackage{booktabs} 
\usepackage{subcaption}
\usepackage{multirow}
\usepackage{textcomp, color, soul}
\usepackage{tabularx}
\usepackage{url}
\usepackage{textcomp, color, soul, xcolor}
\usepackage{color,soul}
\usepackage{colortbl}
\usepackage{amsmath, wrapfig}
\usepackage{ctable}
\usepackage{capt-of}
\usepackage{balance,bm}
\usepackage [english]{babel}
\usepackage{url}
\usepackage [autostyle, english = american]{csquotes}
\usepackage{url}
\usepackage{xcolor}
\usepackage{url}
\usepackage{siunitx}
\usepackage{textcomp, color, soul}
\usepackage{color,soul}
\usepackage{amsmath}
\usepackage{ctable}
\usepackage[show]{chato-notes}

\newcommand{\red}[1]{\textcolor{red}{#1}}

\definecolor{green}{rgb}{0.0, 0.65, 0.31}
\definecolor{bleudefrance}{rgb}{0.19, 0.55, 0.91}
\definecolor{ceruleanblue}{rgb}{0.16, 0.32, 0.75}
\definecolor{grey}{HTML}{969696}
\definecolor{violet}{HTML}{756bb1}
\definecolor{dgrey}{HTML}{404040}
\definecolor{lgrey}{HTML}{bababa}
\definecolor{dgreen}{HTML}{005a32}
\definecolor{purple}{HTML}{ae017e}
\definecolor{neutralCol}{HTML}{dd1c77}
\definecolor{neutralGreen}{HTML}{31a354}
\definecolor{NewBlue}{HTML}{1879ba}
\definecolor{bleudefrance}{rgb}{0.19, 0.55, 0.91}  
\definecolor{AfTrColor}{HTML}{0868ac}  
\definecolor{BfTrColor}{HTML}{a8ddb5}  
\definecolor{AfCtColor}{HTML}{b10026}  
\definecolor{BfCtColor}{HTML}{fd8d3c}
\definecolor{leftcolor}{HTML}{002C77}
\definecolor{rightcolor}{HTML}{CC0C2F}

\definecolor{burntorange}{rgb}{0.8, 0.33, 0.0}
\definecolor{brightmaroon}{rgb}{0.76, 0.13, 0.28}
\definecolor{limegreen}{rgb}{0.0, 0.8, 0.6}
\definecolor{teal}{rgb}{0.0, 0.5, 0.5}
\definecolor{tractorred}{rgb}{0.99, 0.05, 0.21}
\definecolor{falseclaim}{HTML}{d01c8b}
\definecolor{falsecolor}{HTML}{f1b6da}

\DeclareRobustCommand{\hlfc}[1]{{\sethlcolor{falsecolor}\hl{#1}}}

\newcolumntype{P}[1]{>{\centering\arraybackslash}p{#1}}
\newcolumntype{M}[1]{>{\centering\arraybackslash}m{#1}}

\colorlet{tableheadcolor}{gray!25} 
\colorlet{tablerowcolor}{gray!25} 
\colorlet{tablerowcolor2}{gray!12} 
\newcommand{\rowcol}{\rowcolor{tablerowcolor}} %
\newcommand{\rowcollight}{\rowcolor{tablerowcolor2}} %

\setcopyright{acmcopyright}
\acmConference[FAT* '19]{FAT* '19: Conference on Fairness, Accountability, and Transparency}{January 29--31, 2019}{Atlanta, GA, USA}
\acmDOI{10.1145/3287560.3287580}
\acmISBN{978-1-4503-6125-5/19/01}

\begin{document}
\title{On Microtargeting Socially Divisive Ads:\\[0.1em] A Case Study of Russia-Linked Ad Campaigns on Facebook}


\author{Filipe N. Ribeiro\mbox{*}}
\affiliation{%
  \institution{UFOP/UFMG, Brazil}
  }
\email{filipe.ribeiro@ufop.edu.br}

\author{Koustuv Saha\mbox{*}}
\affiliation{%
  \institution{Georgia Tech, US}}
  \email{koustuv.saha@gatech.edu}
  \thanks{\mbox{*} \textbf{\sffamily These authors contributed equally to this work.}}
  
  \author{Mahmoudreza Babaei}
\affiliation{%
  \institution{MPI-SWS, Germany}}
\email{babaei@mpi-sws.org}  
  
\author{Lucas Henrique}
\affiliation{%
  \institution{UFMG, Brazil}}
 \email{lhenriquecl@dcc.ufmg.br}
  
\author{Johnnatan Messias}
\affiliation{%
  \institution{MPI-SWS, Germany}}
\email{johnme@mpi-sws.org}  

\author{Fabricio Benevenuto}
\affiliation{%
  \institution{UFMG, Brazil}}
\email{fabricio@dcc.ufmg.br}  

\author{Oana Goga}
\affiliation{%
  \institution{Univ. Grenoble Alpes, CNRS, Grenoble INP, LIG, France}}
\email{oana.goga@univ-grenoble-alpes.fr}

\author{Krishna P. Gummadi}
\affiliation{%
  \institution{MPI-SWS, Germany}}
\email{gummadi@mpi-sws.org}

\author{Elissa M. Redmiles}
\affiliation{%
  \institution{University of Maryland, US}}
\email{eredmiles@cs.umd.edu}  

\begin{abstract}
Targeted advertising is meant to improve the efficiency of matching advertisers to their customers. However, targeted advertising can also be abused by malicious advertisers to efficiently reach people susceptible to false stories, stoke grievances, and incite social conflict.
Since targeted ads are not seen by non-targeted and non-vulnerable people, malicious ads are likely to go unreported and their effects undetected. This work examines a specific case of malicious advertising, exploring the extent to which political ads~\footnote{We deployed a system that shows the ads
and the demographics of their targeting audiences (available at \textit{\url{http://www.socially-divisive-ads.dcc.ufmg.br/}}).} from the Russian Intelligence Research Agency (IRA) run prior to 2016 U.S. elections exploited Facebook's targeted advertising infrastructure to efficiently target ads on divisive or polarizing topics (e.g., immigration, race-based policing) at vulnerable sub-populations. In particular, we do the following: (a) We conduct U.S. census-representative surveys to characterize how users with different political ideologies \textit{report}, \textit{approve}, and \textit{perceive truth in} the content of the IRA ads.
Our surveys show that many ads are ``divisive'': they elicit very different reactions from people belonging to different socially salient groups. (b) We characterize how these divisive ads are targeted to sub-populations that feel particularly aggrieved by the status quo. 
Our findings support existing calls for greater transparency of content and targeting of political ads. (c) We particularly focus on how the Facebook ad API facilitates such targeting. We show how the enormous amount of personal data Facebook aggregates about users and makes available to advertisers enables such malicious targeting. 

\end{abstract}


\maketitle

\section{Introduction}


Online targeted advertising refers to the ability of an advertiser to select audience for their ads. Such advertising constitutes the primary source of revenue for many online sites including most social media websites such as Facebook, Twitter, YouTube, and Pinterest. Consequently, these websites accumulate detailed demographic, behavioral and interest profiles of their users enabling advertisers to ``microtarget'', i.e., choose small (tens or hundreds to thousands) of users with very specific attributes like people living in a zipcode that read New York Times or Breitbart. Beyond raising numerous privacy concerns~\cite{korolova2011JournalPrivacyandConfidentiality,venkatadri2018privacy}, targeted advertising platforms have come under scrutiny for enabling {\it discriminatory advertising}, where ads announcing housing or job opportunities are targeted to exclude people belonging to certain races or gender~\cite{propublica_fb_ethnic_affinity,fb_ethnic_affinity_ban, speicher-2018-targeted, anupamdatta_CMU_ad_discrimination}. 

In this paper, we analyze the potential for a new form of abuse on targeted advertising platforms namely, {\it socially divisive advertising}, where malicious advertisers incite social conflict by publishing ads on divisive societal issues of the day (e.g., immigration and racial-bias in policing in the lead up to 2016 US presidential elections). Specifically, we focus on how ad targeting on social media sites such as Facebook can be leveraged to selectively target groups on different sides of a divisive issue with (potentially false) messages that are deliberately crafted to stoke their grievances and thereby, worsen social discord. We also investigate whether targeted ad platforms allow such malicious campaigns to be carried out in stealth, by excluding people who are likely to report (i.e., alert site administrators or media watchdog groups about) such ads.



Our study is based on an in-depth analysis of a publicly released dataset of Facebook ads run by a Russian agency called Internet Research Agency (IRA) before and during the American Election on the year of 2016 \footnote{\textit{\url{www.wsj.com/articles/you-cant-buy-the-presidency-for-100-000-1508104629}}} \footnote{\textit{\url{www.nytimes.com/2017/11/01/us/politics/russia-2016-election-facebook.html}}}. Our analysis is centered around three high-level research questions:

\vspace{0.5em}
\noindent {\it RQ 1: How divisive is the content of the IRA ads?} We quantify the divisiveness of an ad by analyzing the {\it differences in reactions} of people with different ideological persuasions to the ad. Specifically, using US census-representative surveys, we look at how conservative- and liberal-minded people differ in (a) how likely they are to report the ad, (b) how strongly they approve or disapprove the ad's content, and (c) how they perceive truthood (or falsehood) in ad's claims. Our analysis shows that IRA ads elicit starkly different and polarizing responses from people with different ideological pursuasions.   

\vspace{0.5em}
\noindent {\it RQ 2: How effectively done was the targeting of the socially divisive ads?} We find that the ``Click Through Rate'' (CTR), a traditional measure of effectiveness of targeting, of the IRA ads are an order of magnitude (10 times) higher than that of typical Facebook ads. The high CTR suggests that the ads have been targeted very efficiently. A deeper analysis of the demographic biases in the targeted audience reveals that the ads have been targeted at people who are more likely to approve the content and perceive fewer false claims, and are less likely to report.   

\vspace{0.5em}
\noindent {\it RQ 3: What features of Facebook's ad API were leveraged in targeting the ads?} We also analyze the construction or specification of ``targeting formulae'' for the ads, i.e., the combination of Facebook user attributes that are used when selecting the audience for the ads. We find widespread use of interest attributes such as ``Black Consciousness movement'' and ``Chicano movement'' that are mostly shared by people from specific demographic groups such as African-Americans and Mexican-Americans. We show how Facebook ad API's suggestion feature may be exploited by the advertisers to find interest attributes that correlate very strongly to specific social demographic groups.

\subsection{Related Work}

Prior work has highlighted several forms of abuses of targeted advertising in Facebook, such as for inappropriately exposing the private information of users to advertisers~\cite{venkatadri2018privacy}, and for allowing discriminatory advertising (e.g., to exclude users belonging to a certain race or gender from receiving their ads)~\cite{speicher-2018-targeted}. Our effort highlights a new and different form of potential abuse of these targeted advertising platforms in creating a social discord.

A rich body of prior work have focused on understanding filter bubbles, echo chambers, polarization, and ideological discourse in social media as an emergent phenomenon~\cite{castillo2014characterizing,del2016echo,flaxman2016filter,garimella2018political,guerra2013measure,Lima2018@asonam,sharma2017analyzing}. We provide a complementary perspective on the topic by examining how echo chambers and polarization can be engineered on social media through targeted advertising. 
A recent work conducted a detailed study about Facebook Ads environment by analyzing thousands of ads collected through a browser plugin\cite{Andreou19a}. More closely related to our work,~\citeauthor{mie2018}
gathered Facebook ads from individuals and analyzed who are behind divisive ad campaigns, reporting 
suspicious foreign entities~\cite{mie2018}. Differently, we focus on understanding the disruptive ability of microtargeting for providing divisive political ad campaigns.

Finally, our effort is complementary to prior work that attempts to understand the abuse of social media by misinformation campaigns, especially along political elections~\cite{lazer2018science, vosoughi2018spread}. Our work provides a better comprehension about a key dissemination mechanism of fake news stories, highlighting how advertising platforms allow injection of misinformation in social systems and choose vulnerable people as the target.

%

\section{Russia-Linked Facebook Ads Dataset}



On May 10th, 2018 the Democrats Permanent Select Committee on Intelligence released a dataset containing 3,517 Facebook advertisements\footnote{\textit{\url{democrats-intelligence.house.gov/facebook-ads/social-media-advertisements.htm}}} from 2015, 2016, and 2017 that are linked to a Russian propaganda group: Internet Research Agency (IRA).

\begin{figure}[t!]
\centering
\includegraphics[scale=0.35]{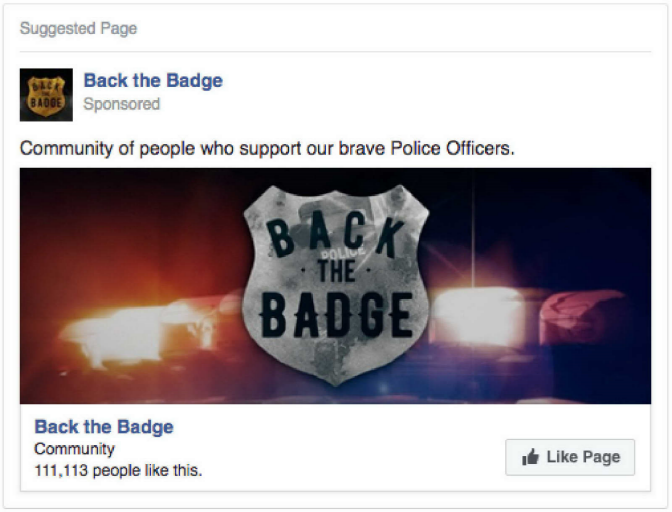}
\caption{\textbf{Example of an Ad from the Dataset.}}
\label{fig:exampleAd}
\end{figure}

Each ad is composed of an image and text
(Figure~\ref{fig:exampleAd} shows an example). 
Additionally, each ad contains a landing page, which is a link to the host of the ad, as well as an ad ID; an ad targeting formula, which is a combination of demographic, behavioral and user interest aspects used to target Facebook users; the cost for running the ad in Russia Rubles\footnote{We converted currency of the costs to USD as of May 15th, $1$ USD = $61.33$ RUB.}; the number of impressions, which is the number of users who spent some time observing the ad; the number of clicks received by the ad; and, finally, the ad creation and end dates. 
This section provides an overview of these ads.

\begin{figure}[!t]
    \centering
    \includegraphics[width=1\columnwidth]{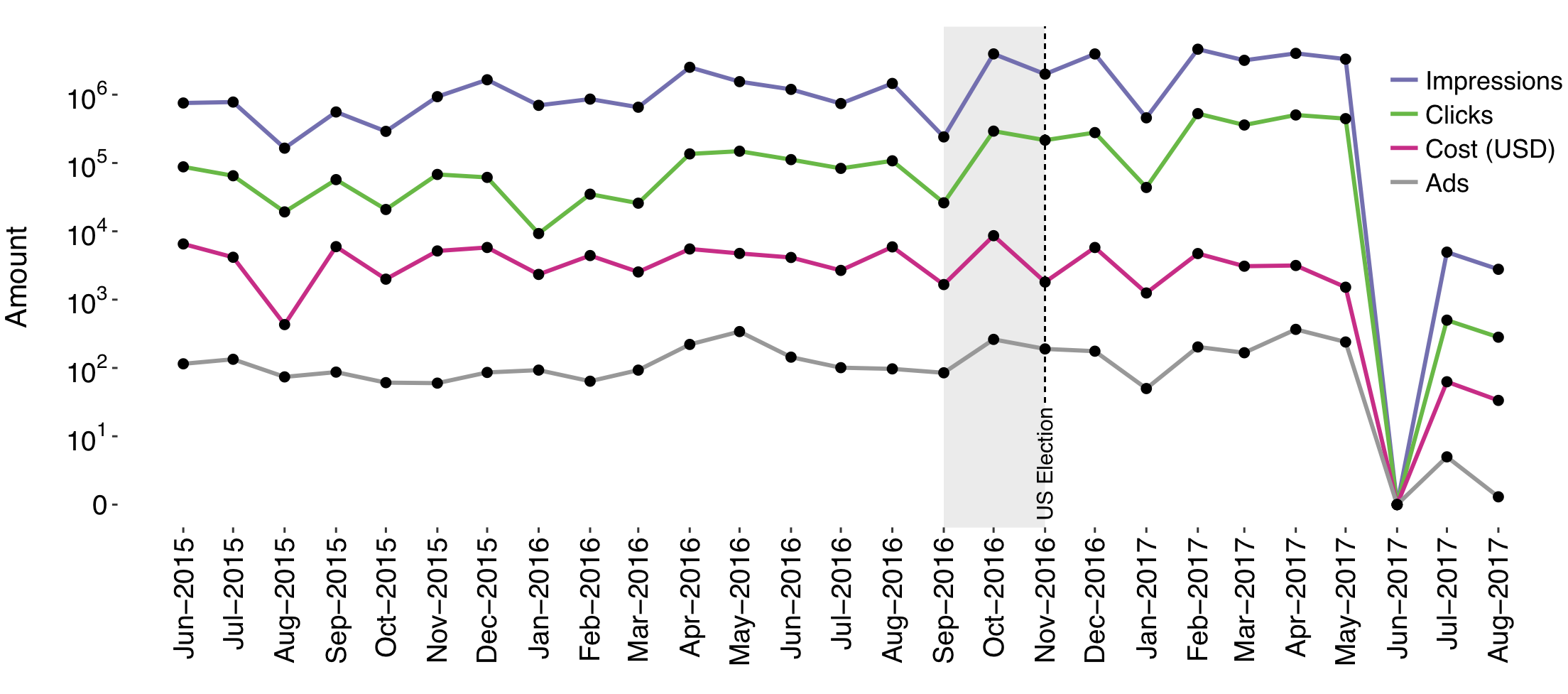}
    \caption{\textbf{Number of ads created, their impressions, cost, and received clicks over time. Shaded region shows the 2-month period just before the 2016 U.S. Election.}}
    \label{fig:temporalAnalysis}
\end{figure}

The ads in the dataset were run between June 2015 and  August 2017. 
From the 3,517 advertisements, we found that 617 ($17.5\%$) were created in 2015, 1,867 ($53.1\%$) in 2016, and 1,033 ($29.4\%$) in 2017.
Figure~\ref{fig:temporalAnalysis} shows the distribution of these ads over time in terms of the number of ads created per month, cost to run the ads, and impressions and clicks received. Note that the y-axis is in log scale. We observed that the number of impressions, and clicks, increases almost an order of magnitude around the election period (shaded region). There is also another peak in February, just after the newly elected U.S. President Donald Trump assumed office.

\subsection{Landing Pages}

\begin{figure}[tb]
    \centering
    \includegraphics[width=1\columnwidth]{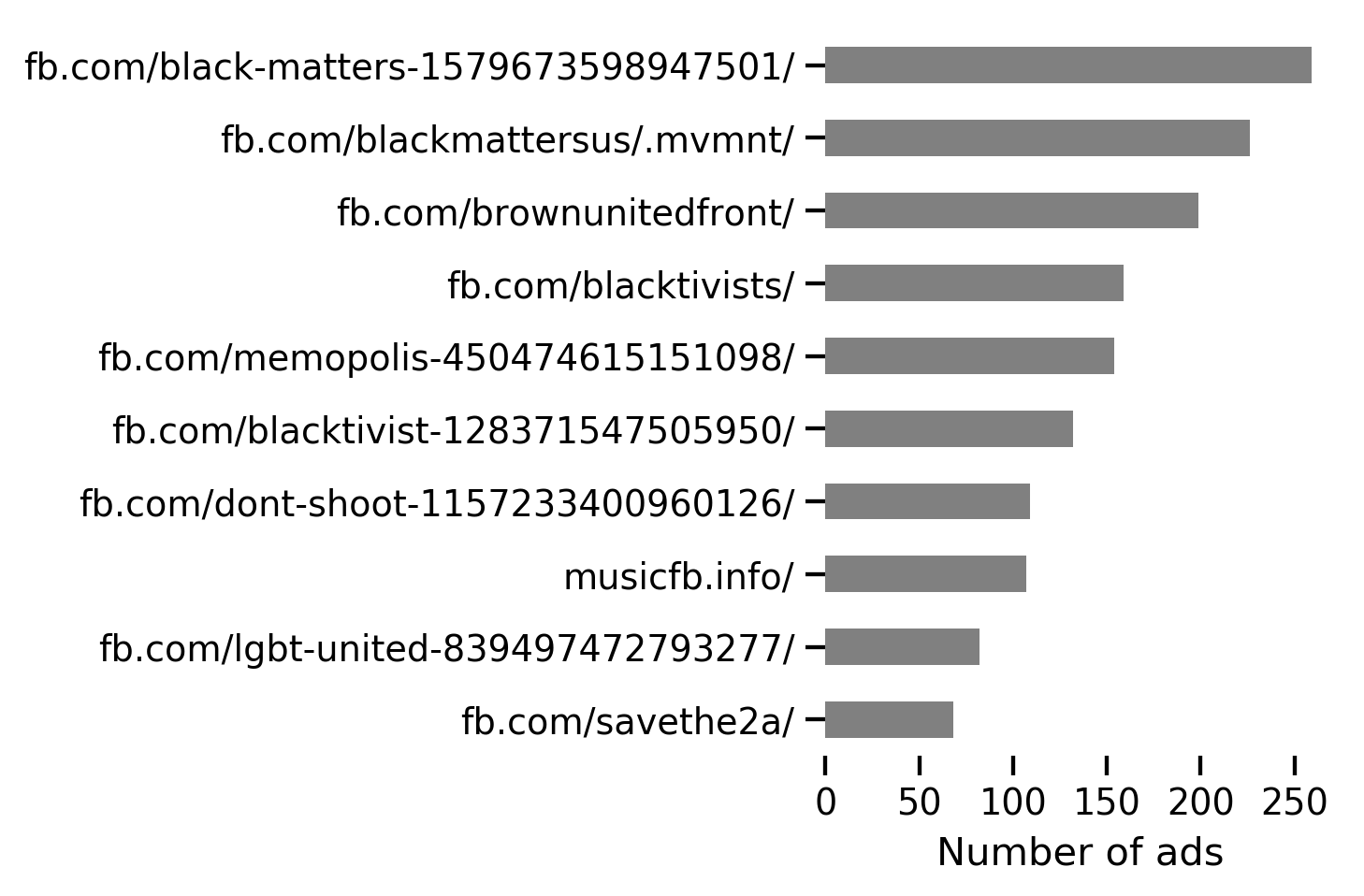}
    \caption{\textbf{Top 10 Landing Pages based on the number of ads.}}
    \label{fig:topLandingPages}
\end{figure}

We first explore the ad landing pages: the \textit{urls} to which users who clicked on the ads were redirected. There are $462$ unique landing pages corresponding to all the ads. Figure~\ref{fig:topLandingPages} shows the top $10$ landing pages per number of ads posted. The most popular landing page (\textit{\url{fb.com/Black-Matters-1579673598947501/}}) posted $259$ advertisements. Interestingly, one of the top landing pages, the \textit{musicfb.info}\footnote{\textit{\url{web.archive.org/web/20161019155736/musicfb.info/}}}, invites users to install a browser extension, which was reported to send spam to the Facebook friends of those who installed it\footnote{\textit{\url{wired.com/story/russia-facebook-ads-sketchy-chrome-extension/}}}. This landing page received 24,623 impressions, $85$ clicks, and spent around US\$$112.38$. The domain \textit{musicfb.info} was also promoted by other pages, accounting for $3\%$ of all ads. We also find that the most popular landing pages are Facebook pages, accounting for $84\%$ of all ads, followed by \textit{\url{blackmattersus.com}} ($7\%$), and Instagram ($3.4\%$). For 28 ads, we were not able to identify their landing pages because these pages were already blocked.
%

%
%



\subsection{Cost, Impressions, Clicks and CTR}
\begin{figure}[t]
 \begin{subfigure}[b]{0.50\columnwidth}
   \centering
  \includegraphics[width=\columnwidth]{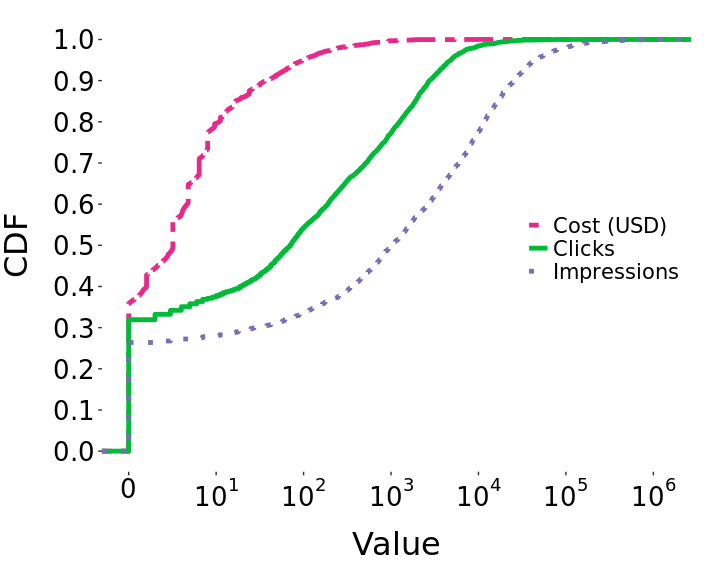}
 \caption{}
 \end{subfigure}\hfill
   \begin{subfigure}[b]{0.50\columnwidth}
   \centering
  \includegraphics[width=\columnwidth]{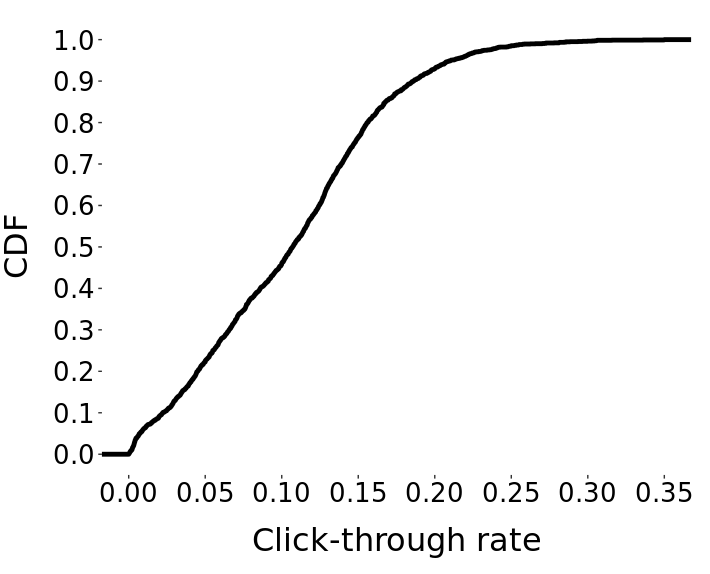}
 \caption{}
   \end{subfigure}\hfill
    \caption{Cumulative Distribution Function (CDF) of the ads on their (a) clicks, impressions, and costs, (b) click-through-rates.}
    \label{fig:CDFsMetrics}
\end{figure}

Figure~\ref{fig:CDFsMetrics} (a) shows the cumulative distribution functions (CDFs) of all the ads in the dataset on their number of impressions, clicks, and amount spent to advertise.
The most expensive ad cost 5,307 \$ USD. The highest number of impressions generated was 1,335,000 and the maximum number of clicks was 73,060.  
Nearly $25\%$ of the landing pages spent more than $100$ dollars, $26.8\%$ of the pages received more than 1,000 clicks, and around $36.1\%$ had more than $10,000$ impressions. On the other hand, more than $25\%$ of the ads had no impressions, clicks, and cost, suggesting these ads were not launched or ran for a very short period of time.

An average ad cost $34.5$ USD, was seen by 11,536 users, and received 1,062 clicks. 
The average value is increased to $38$ USD for cost, 16,482 for impressions, and 1,521 for the number of clicks if we exclude those ads that appeared not to have been run. The Pearson's correlation coefficient among cost, impressions, and clicks is very high, particularly between impressions and clicks ($0.89$). We also noted that this dataset is quite skewed, as $10\%$ of the ads accumulate $85.18\%$ of the total cost, $71.93\%$ of the total number of impressions, $69.47\%$ of the total number of clicks. 

However, there were notable exceptions to this correlation: higher investment (cost) did not always lead to higher return (e.g., impressions, clicks). Table~\ref{table:topLandingPages} shows the most popular landing pages per impressions, clicks, and cost of the ads. For example, \url{fb.com/brownunitedfront/}, received the largest number of impressions (5,817,734), corresponding alone to $14.3\%$ of impressions obtained by all ads, but cost only $6.5\%$ of the total cost of all ads in the dataset.

\begin{table*}[tb]
\centering
\small
\begin{tabular}{lrlrlr}
\hline
\multicolumn{2}{c}{Impressions}& \multicolumn{2}{c}{Clicks}& \multicolumn{2}{c}{Cost (USD)}\\ 
\hline
fb.com/brownunitedfront/& 14.3\%&  fb.com/brownunitedfront/ & 18.8\%& fb.com/patriototus/   & 6.5\% \\ 
\rowcollight fb.com/blacktivists/  & 10.8\% & fb.com/Blacktivist-128371547505950/ & 13.8\%  &  fb.com/blacktivists/                       & 5.4\%                             \\ 
fb.com/Blacktivist-128371547505950/ & 10.5\%  & fb.com/blacktivists/ & 11.9\%& fb.com/blackmattersus/& 5.3\%\\ 
\rowcollight fb.com/blackmattersus.mvmnt/                & 4.7\%                     & fb.com/blackmattersus.mvmnt/               & 7.0\%& fb.com/timetosecede/                       & 4.7\%                             \\ 
fb.com/Woke-Blacks-294234600956431/         & 3.3\%                     & fb.com/Dont-Shoot-1157233400960126/        & 3.6\% & fb.com/Igbtun/                             & 4.3\%                            \\ 
\rowcollight fb.com/copsareheroes/                       & 3.3\%    & fb.com/blackmattersus/                     & 2.5\%& fb.com/BlackJourney2Justice/               & 4.1\%                             \\ 
fb.com/blackmattersus/                      & 3.1\%         & fb.com/patriototus/                        & 2.5\%  & fb.com/MuslimAmerica/& 3.28\\ 
\rowcollight fb.com/South-United-1777037362551238/& 2.7\%& fb.com/Memopolis-450474615151098/& 2.4\%& fb.com/South-United-1777037362551238/& 3.2\%\\ 
fb.com/Dont-Shoot-1157233400960126/& 2.2\%& fb.com/Woke-Blacks-294234600956431/& 2.3\%&fb.com/blackmattersus.mvmnt/& 2.7\%\\ 
\rowcollight fb.com/patriototus/& 1.7\%& fb.com/South-United-1777037362551238/& 2.0\%& fb.com/savethe2a/& 2.5\%\\ 
\bottomrule
\end{tabular}
\caption{\textbf{Most popular landing pages per impressions, clicks, and cost. }}
\label{table:topLandingPages}
\end{table*}

Finally, we compute the click-through rate (CTR) of these ads, which is a typical metric to measure the effectiveness of an ad. It is computed as a ratio between the number of clicks and the number of impressions received by an ad. 
Figure~\ref{fig:CDFsMetrics} (right) shows the cumulative distribution function of the CTR of the ads, excluding those with $0$ values for clicks, impressions, and cost. The median CTR is $10.8\%$ and $75\%$ of the ads have a CTR higher than $5.6$. The average CTR is $10.8\%$. These are incredibly high values for CTR. As a comparison, WordStream released a report as of April 2018\footnote{\textit{\url{wordstream.com/blog/ws/2017/02/28/facebook-advertising-benchmarks}}} which shows the average CTR for Facebook ads across all industries is $0.9\%$. As an example, Retail is $1.6\%$, Fitness is $1\%$, Health care $0.8\%$, and Finance is $0.56\%$. This means that these political ads have a CTR that is about an order of magnitude higher than a typical Facebook ad.  

\subsection{High Impact Ads}
Our analysis reveals 	 that only a few ads are responsible for most of the cost, impressions, and clicks. Considering this, we defined a set of high impact ads as the union of the top $10\%$ ads in terms of cost, impressions, clicks, and CTR. We obtained $905$ high impact ads, corresponding to $27.7\%$ of the entire dataset. These ads account together to $83.9\%$ of the total number of impressions, $81.8\%$ of clicks, $88.5\%$ of the cost, and $46.9\%$ of the CTR. 
For the purposes of our study, where we require manual inspection of the ads (to identify their targets and to run surveys), our ensuing analyses concern those high impact ads run before the 2016 U.S. elections: $485$ ads.

\subsection{Summary}

This section describes and characterizes the ads in the IRA dataset. Our analysis highlights the landing pages that paid for the ads and identifies the most successful ads in terms of impressions and clicks. We find that the ad campaigns were intensified near to the U.S. election period. Among our main findings, we show that the typical CTR for these ads is an order of magnitude higher than typical values for Facebook, meaning that these ads were very effective. 

\section{Analyzing Divisiveness of the Ads}

To investigate whether these ads were designed to be ideologically divisive -- that is, designed to elicit different reactions from people with different political viewpoints -- we conducted three online surveys on a U.S. census-representative sample (n=2,886). 
We used each survey to measure one of three axes along which ads could potentially be divisive:  1) \textit{reporting}: whether respondents would report the ads, and why, 2) \textit{approval and disapproval}: whether they approve or disapprove the content of the ad, and 3) \textit{false claims}: if they are able to identify any false claims in the content of the ad.

Our surveys considered only those 485 \textit{high impact} ads which were run before the elections. 
Each survey showed
ten ads followed by demographic questions. More detail on the specific questions used to assess each axis is provided in the corresponding axis subsections that follow. The survey questions were pre-tested using cognitive interviews and all survey questions included a ``I don't know'' or ``Prefer not to respond'' answer choice to ensure internal measurement validity~\cite{beatty2007research}. 
To obtain a demographically representative sample, and ensure that we captured a wide variety of American perceptions, we deployed the surveys using the Survey Sampling International survey panel\footnote{https://www.surveysampling.com/audiences/consumer-online/}, a non-probabilistic census-representative survey panel. For each survey, we sampled at least 730 respondents (15 responses per ad) whose demographics were representative of the U.S. within 5\% and who had a range of political views (40\% liberal, 40\% conservative, and 20\% moderate or neutral); across the three surveys we obtained a total sample of 2,886 respondents.





We measured overall ideological divisiveness on the three axes (reporting, approval, and false claims) using two metrics:

\noindent\textit{\textbf{Within-group divisiveness.}} Within-group divisiveness measures the extent to which respondents' answers about a particular ad are consistent with their political ideology. That is, do all liberals answer similarly about a particular ad. For each ad, we first calculate the standard deviation of \textit{all} the responses, and then we calculate the standard deviation of the responses within a particular ideological group. Next, we compute within-group divisiveness as the fraction of within-group standard deviation to the overall standard deviation.
Therefore we interpret values lower than 1 as lower divisiveness (and greater agreeableness) within a group than overall, and values greater than 1 as greater within-group divisiveness than overall.




\noindent\textit{\textbf{Between-group divisiveness. }} 
Between-group divisiveness measures the extent to which answers from respondents of one political ideology differ from answers of respondents who align with another political ideology. That is, do liberals answer differently about a particular ad than conservatives. For an ad, we calculate the difference between the mean responses per ideological group, and then compute the fraction of this difference over the maximum possible difference given the range of values to obtain the between-group divisiveness measure. This limits the range of between-group divisiveness measure between 0 and 1, where higher values indicate greater divisiveness between ideological groups.

Table~\ref{table:divisivenessSummary} summarizes the divisiveness of the high impact ads. We find that the within-group divisiveness measure is lower than 1 for all our surveys. This indicates high agreeableness within the ideological groups. In addition, about 20\% of the ads show between-group divisiveness higher than 0.5, indicating severe divisiveness between ideological groups for those ads.

\begin{table}[t]
\centering
\small
    \begin{tabular}{lrrrrrr}
    \toprule
 Measure (Group) & \multicolumn{2}{c} {Reporting}&  \multicolumn{2}{c} {Approval}& \multicolumn{2}{c} {False Claims}\\
 {} & Mean & Stdev. & Mean & Stdev. & Mean & Stdev.\\
 \toprule
     \rowcol \multicolumn{7}{l}{\textit{Within-group divisiveness}}\\
     Liberals & 0.87 &  0.47 & 0.92 & 0.36 & 0.66& 0.69\\
     Conservatives &  0.90 & 0.43 & 0.98 & 0.31 & 0.86& 0.63\\
     \rowcol \multicolumn{7}{l}{\textit{Between-group divisiveness}}\\
     Political & 0.24 & 0.18 & 0.34 & 0.24 & 0.17 & 0.14\\
\bottomrule
\end{tabular}
\caption{Divisiveness measures of the high impact ads.}
\label{table:divisivenessSummary}
\end{table}




\subsection{Likelihood of reporting the ads}

The first axis of divisiveness that we explored was reporting. We surveyed respondents regarding:  
1) Whether they would report the ad shown?\footnote{Specifically, we asked ``Some social media platforms allow you to report content by clicking "report". Would you report this ad (e.g., Mark it as inappropriate or offensive)'' With answer choices ``Yes'', ``No'', ``I don't know''.}, and 2) If they would, why do they find the ad inappropriate? Answer choices given, drawn directly from Facebook's reporting interface~\cite{fb:report}, were: \textit{sexually inappropriate}, \textit{violent}, \textit{offensive}, \textit{misleading}, \textit{disagree}, \textit{false news}, \textit{spam}, and \textit{something else}. 

\begin{figure}[t!]
\centering
    \begin{subfigure}[b]{0.5\columnwidth}
    \centering
  \includegraphics[width=\columnwidth]{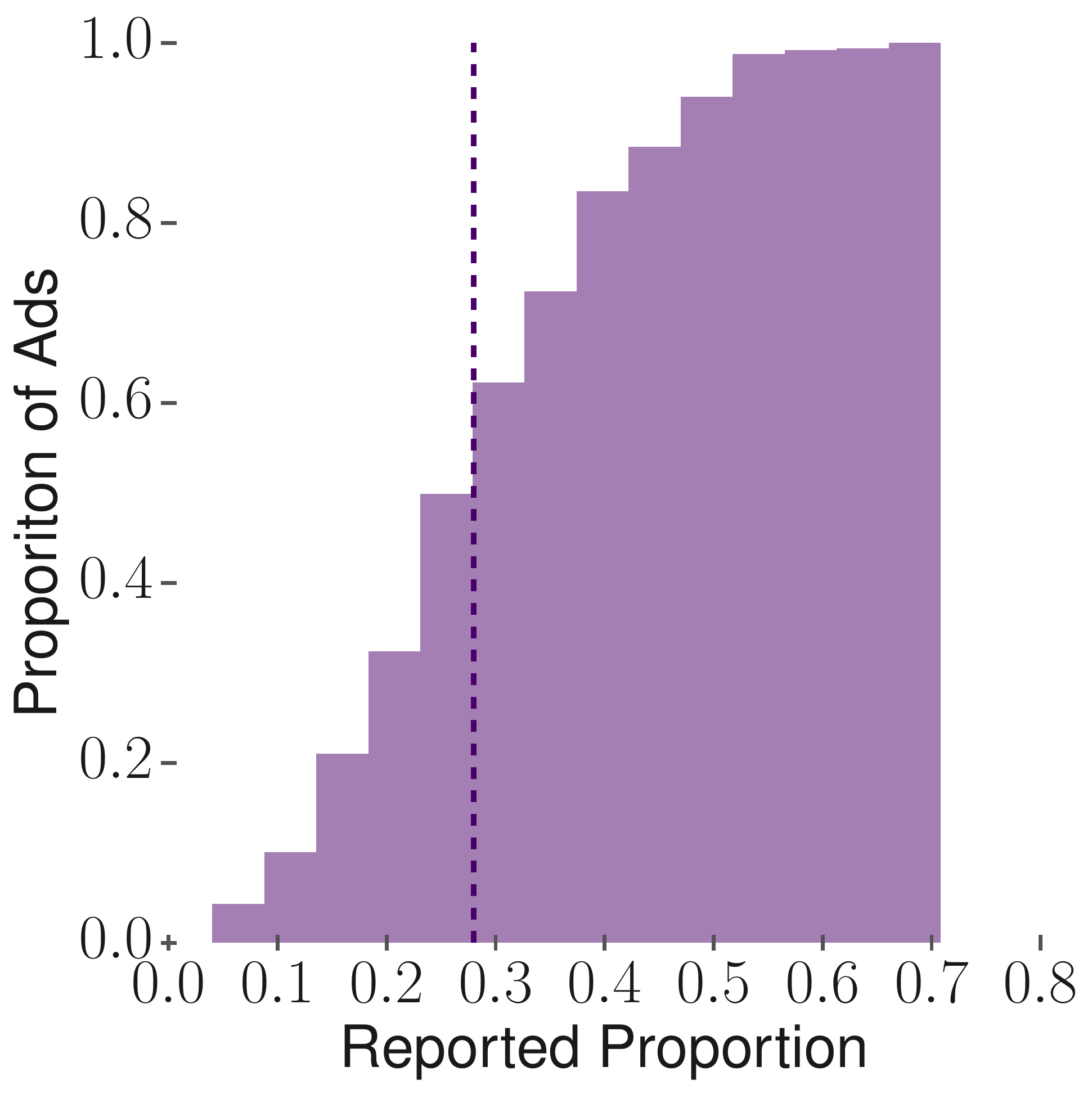}
    \caption{Reported Ads}
    \end{subfigure}\hfill
\begin{subfigure}[b]{0.5\columnwidth}
    \centering
  \includegraphics[width=\columnwidth]{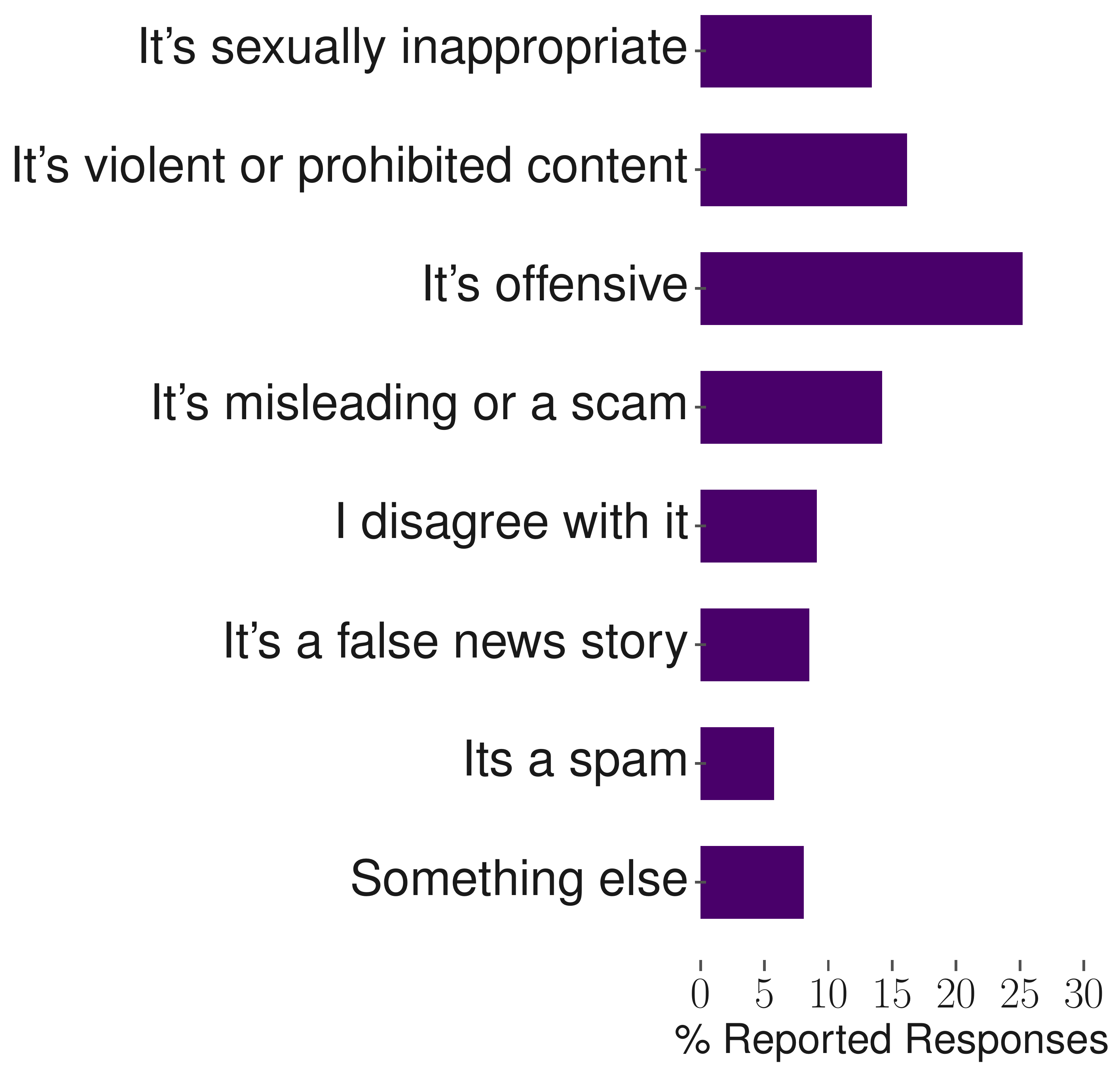}
    \caption{Reasons of Inappropriateness}
    \end{subfigure}\hfill
     \caption{Distribution of the high impact ads on the (a) proportion of reported ads in our dataset, (b) reasons of inappropriateness.}
    \label{fig:reportingComparison}
\end{figure}

Figure~\ref{fig:reportingComparison} shows the reporting responses for the high impact IRA ads. 
For over 73\% of these ads, at least 20\% of the respondents responded that they would have reported the ads. We observe that the majority of the ads were reported on the grounds of being offensive (25\%), violent (15\%), and misleading (15\%).  
Additionally, a substantial proportion (9\%) of the reported responses belonged to the \textit{something else} category. 
In such cases, the respondents entered free-text to explain their reason for inappropriateness. Out of the 61 responses that we received in the free-text box, the pre-dominant reasons were that the ad incites racism (20\%), and that the ad creates divide (5\%) in the society.

Next, to examine ideological divisiveness, we find that the mean within-group divisiveness is 0.87 (stdev = 0.47) for liberals and 0.90 (stdev = 0.43) for conservatives. Both of these within-group divisiveness measures being less than 1, suggests that the likelihood with which individuals within the same ideological group agree about reporting an ad is higher than that when compared against individuals across ideological groups.

\begin{figure}[t!]
\centering

    \begin{subfigure}[b]{0.50\columnwidth}
    \centering
  \includegraphics[width=\columnwidth]{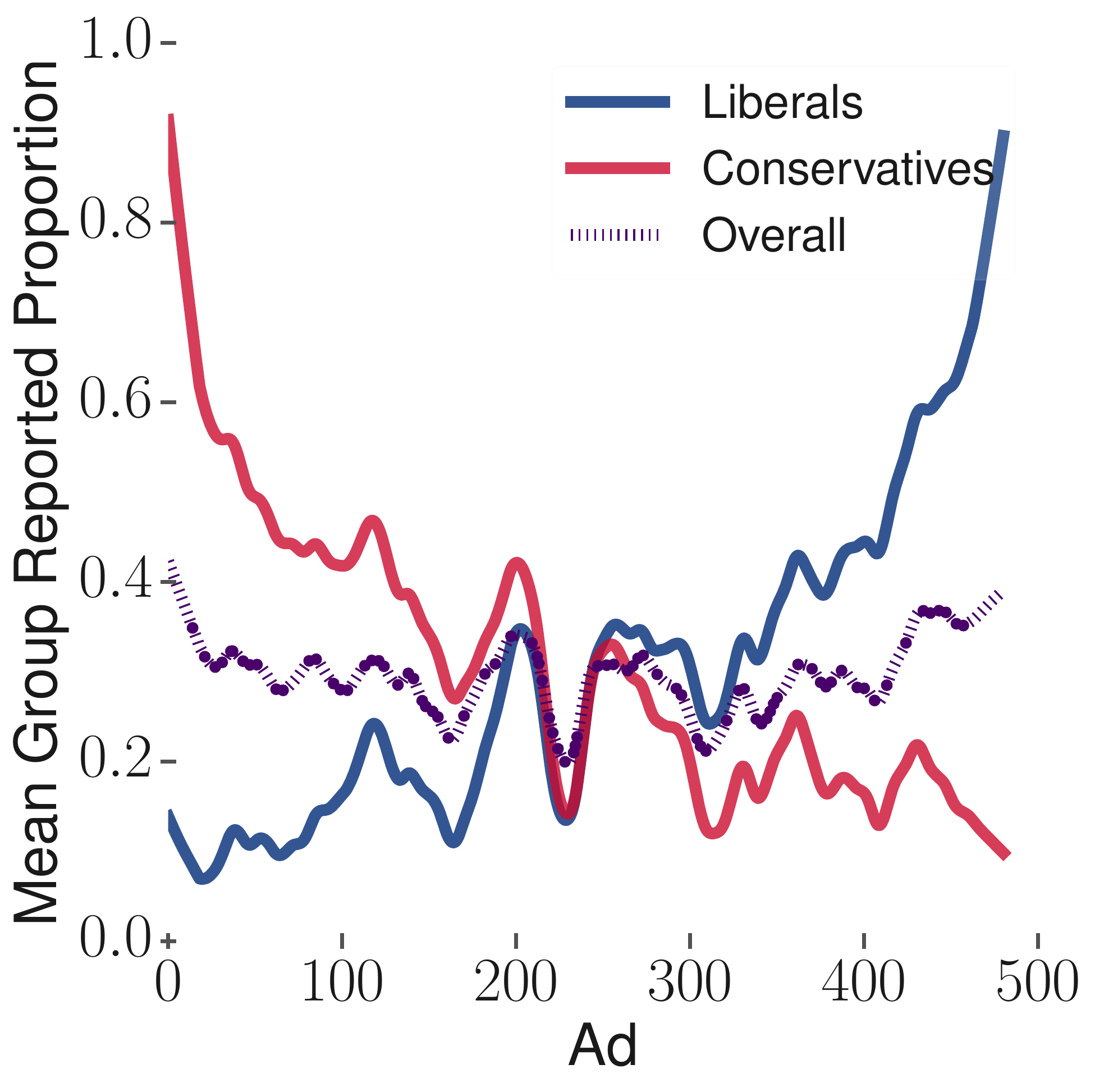}
    \caption{Ideological differences}
    \end{subfigure}\hfill
      \begin{subfigure}[b]{0.50\columnwidth}
    \centering
  \includegraphics[width=\columnwidth]{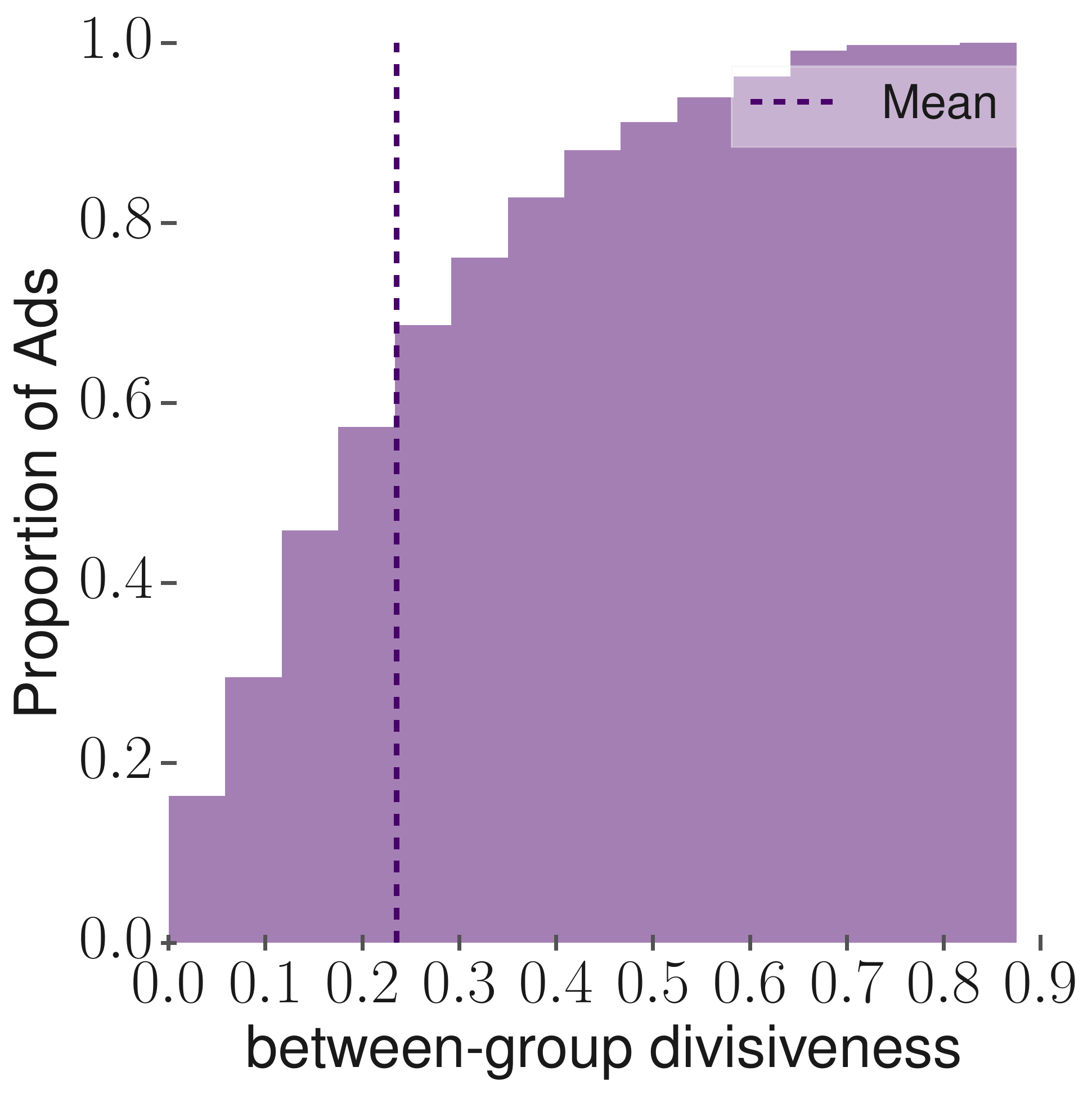}
    \caption{Between-group divisiveness}
    \end{subfigure}\hfill
    \caption{Distribution of reporting across ideological groups. (a) shows the distribution o proportion of the ads being reported by either of political ideology, with x-axis containing each of the high impact ads, (b) plots the between-group divisiveness for the high impact ads.}
    \label{fig:reportingContent}
\end{figure}

Figure~\ref{fig:reportingContent} (a, b) shows the distribution of the reporting across ideological groups. We find significant differences in terms of the reporting behavior across political ideologies. Defining a median threshold for divisiveness, we find that in over 50 percent of the ads, liberals and conservatives completely disagreed with each other (eg. conservatives showed \textit{more} than their median reported proportion and liberals showed \textit{less} than their median reported proportion, and the vice versa). Table~\ref{table:reportExamples} shows a few examples of the ads which showed the greatest differences in the reporting behavior by the respondents of two political ideologies. These ads typically mention politically-charged topics. For example, immigration ---  ``TAG YOUR PHOTOS WITH \#TXagainst Send us the reason why don't you want illegals in Texas. Comments, photos, and videos are welcomed!'' --- in this case, presenting a viewpoint associated with the Republican Party, Or police brutality --- ``Police are beyond out of control, help us make this viral! Follow our account in order to spread the truth!'' --- in this case, presenting a viewpoint associated with the democratic party.

\begin{table}[t!]
\renewcommand{\arraystretch}{1.1}
\centering
\tiny
	\begin{tabular}{@{}p{\columnwidth}@{}}
\toprule
Reported by both liberals and  conservatives\\
\hline
\textit{TAG YOUR PHOTOS WITH \#TXagainst Send us the reason why don't you want illegals in Texas. Comments, photos, and videos are welcomed!}\\
\rowcollight \textit{Counter-protest against 'White Power' Confederate rally at Stone Mountain Not My Heritage}\\
\textit{W. Wilson, 36, was found dead on Easter Weekend at the LAPD's detention center in jail cell. According to ABC7 report, the black woman, Wakiesha Wilson, had a disagreement with officers before she was found died. Wilson spoke to her lovely family that v Black Woman Found Dead In Jail Cell After Arguing With Detention Officers I Black Matters Black Matters}\\
\rowcollight \textit{Police are beyond out of control, help us make this viral! Follow our account in order to spread the truth!}\\
\textit{Everything you wanted to know about Clinton's dark side. Clinton FRAUDation}\\
\hline
Reported predominantly by  liberals.\\
\hline
    \textit{Join us to learn more! Why aren't white hoods and white supremacist propaganda illegal here in America? Why are Germans ashamed of their bigotry, while America is proud of it? Black America( @black Blacklivessss}\\
\rowcollight \textit{We simply can't allow Muslims to wear burga, otherwise everybody who wants to commit a crime or terror attack would wear this ugly rug and hide his or hers identity behing it. The risk is too highl Burga and other face covering cloth should be banned from wearing in public!}\\
\textit{Five police officers were killed in an organized attack during the protest in Dallas this Blue Lives Matter}\\
\rowcollight \textit{Black intelligence is one of the most highly feared things in this country.}\\
\textit{Parasite is an organism that lives in or on another organism and benefits by deriving nutrients at the host's expense. About 20 million parasites live in the United States illegally. They exploit Americans and give nothing in return. Isn't it time to get rid of parasites that are destroying our country?}\\
\hline
Reported predominantly by  conservatives.\\
\hline
\textit{Come and march with us on 16 April. Stand with Baltimore. Let's make change! Freddie Gray Anniversary March}\\
\rowcollight \textit{Click Watch More to join us! Let's fight against police brutality together! donotshoot.us Donotshoot.us Don't Shoot}\\
\textit{The USA is exactly the place where cops can't care less about people's civil rights. They are cynical toward the rule of law and disrespectful of the rights of fellow citizens. Details: http://donotshoot.us/}\\
\rowcollight \textit{We Muslims of the United States are subject to Islamophobia from the media where regularly STOP SCAPEGOATING MUSLIMS!}\\
\textit{People, our race is in danger! Together we are an invincible power. Just say your word! Join us! Black Pride}\\
    \bottomrule
\end{tabular}
    \caption{Example ads on the basis of reporting behavior by the respondents from two political ideologies.}
\label{table:reportExamples}
\end{table}

\subsection{Approving content of the ads}

\begin{figure}[t]
\centering
    \begin{subfigure}[b]{0.50\columnwidth}
    \centering
  \includegraphics[width=\columnwidth] 
{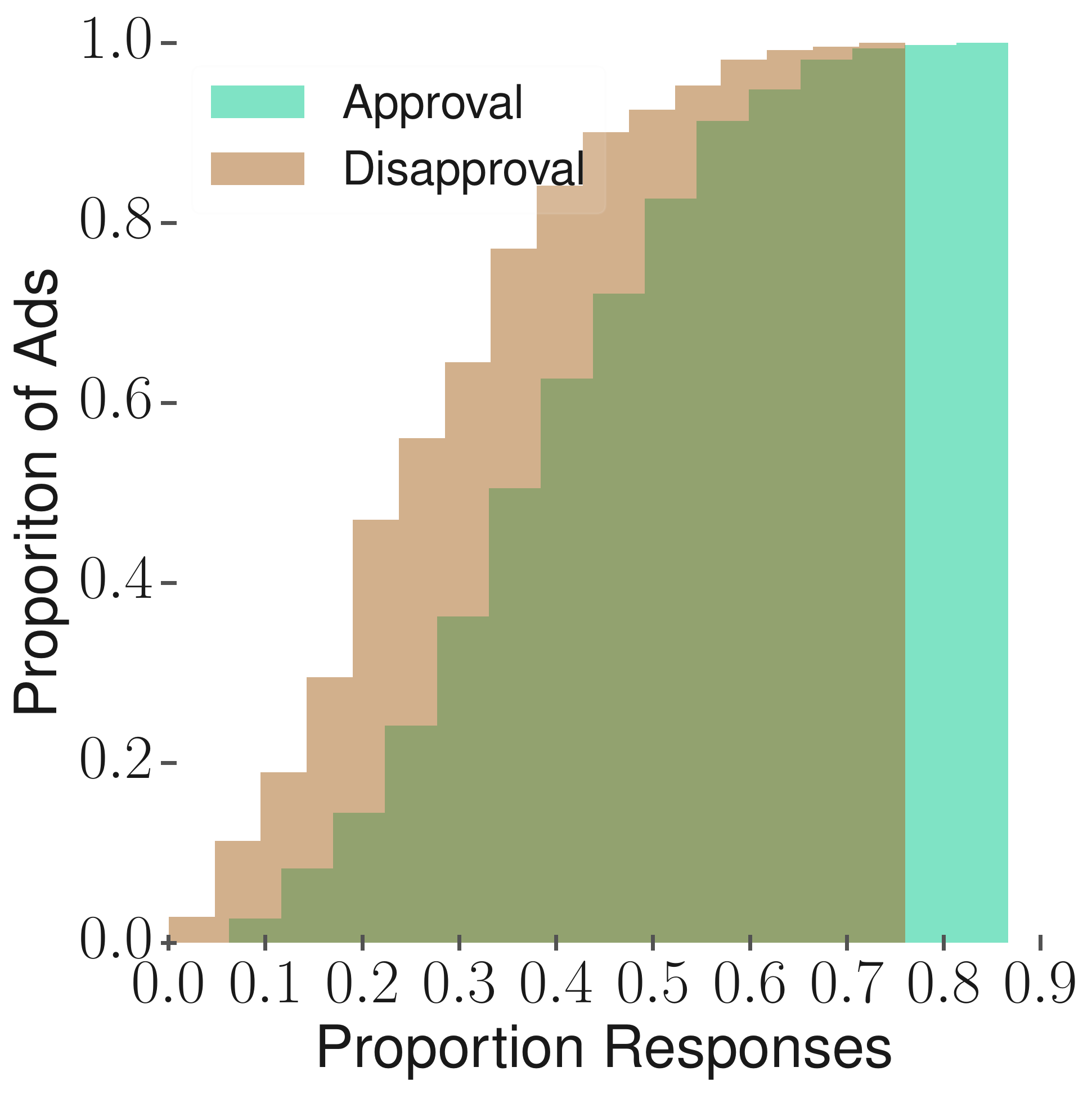}
    \caption{CDF: Proportion Responses}
    \end{subfigure}\hfill
    \begin{subfigure}[b]{0.50\columnwidth}
    \centering
  \includegraphics[width=\columnwidth]{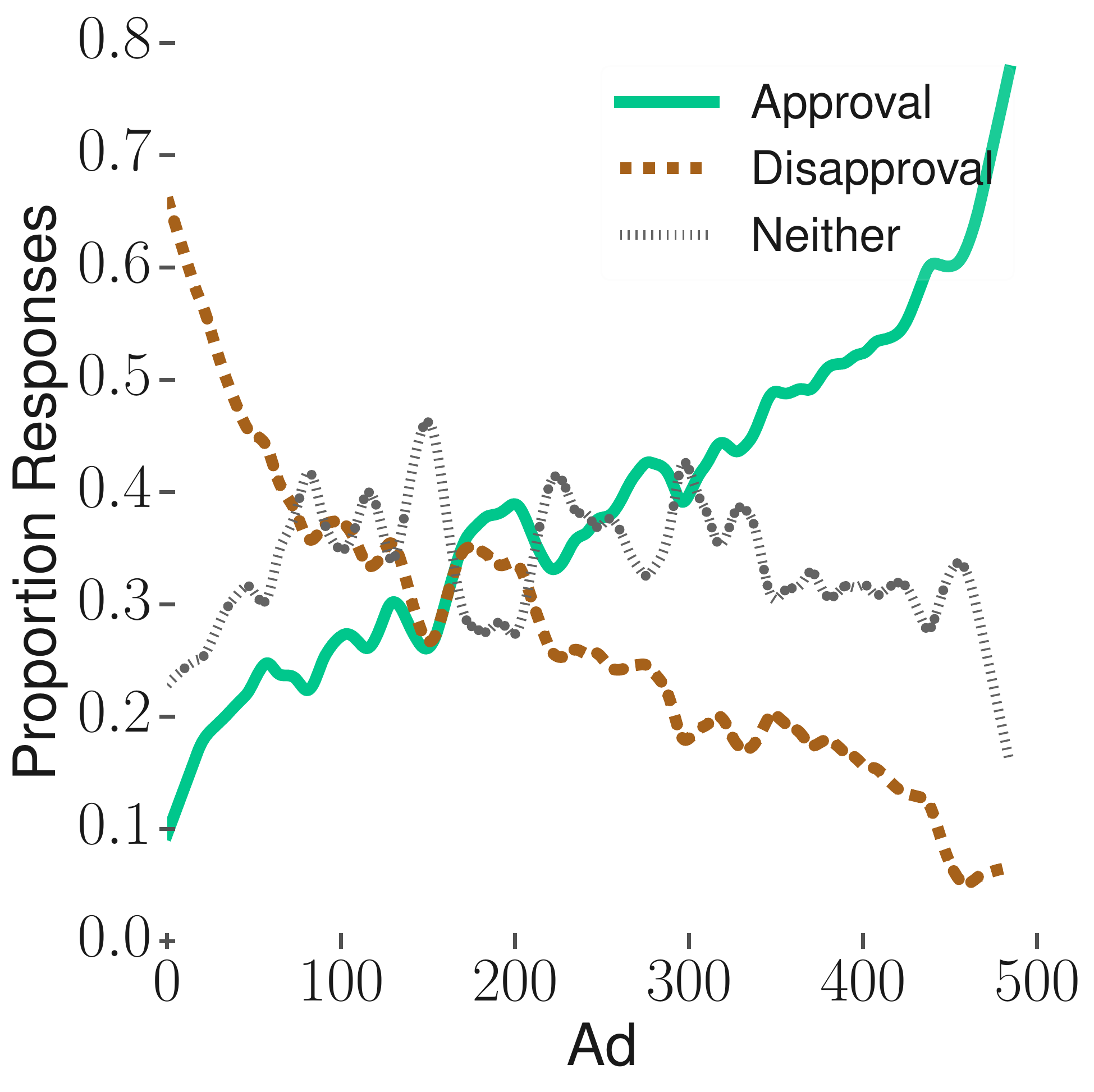}
    \caption{Approval per ad}
    \end{subfigure}
    \begin{subfigure}[b]{0.50\columnwidth}
    \centering
  \includegraphics[width=\columnwidth]{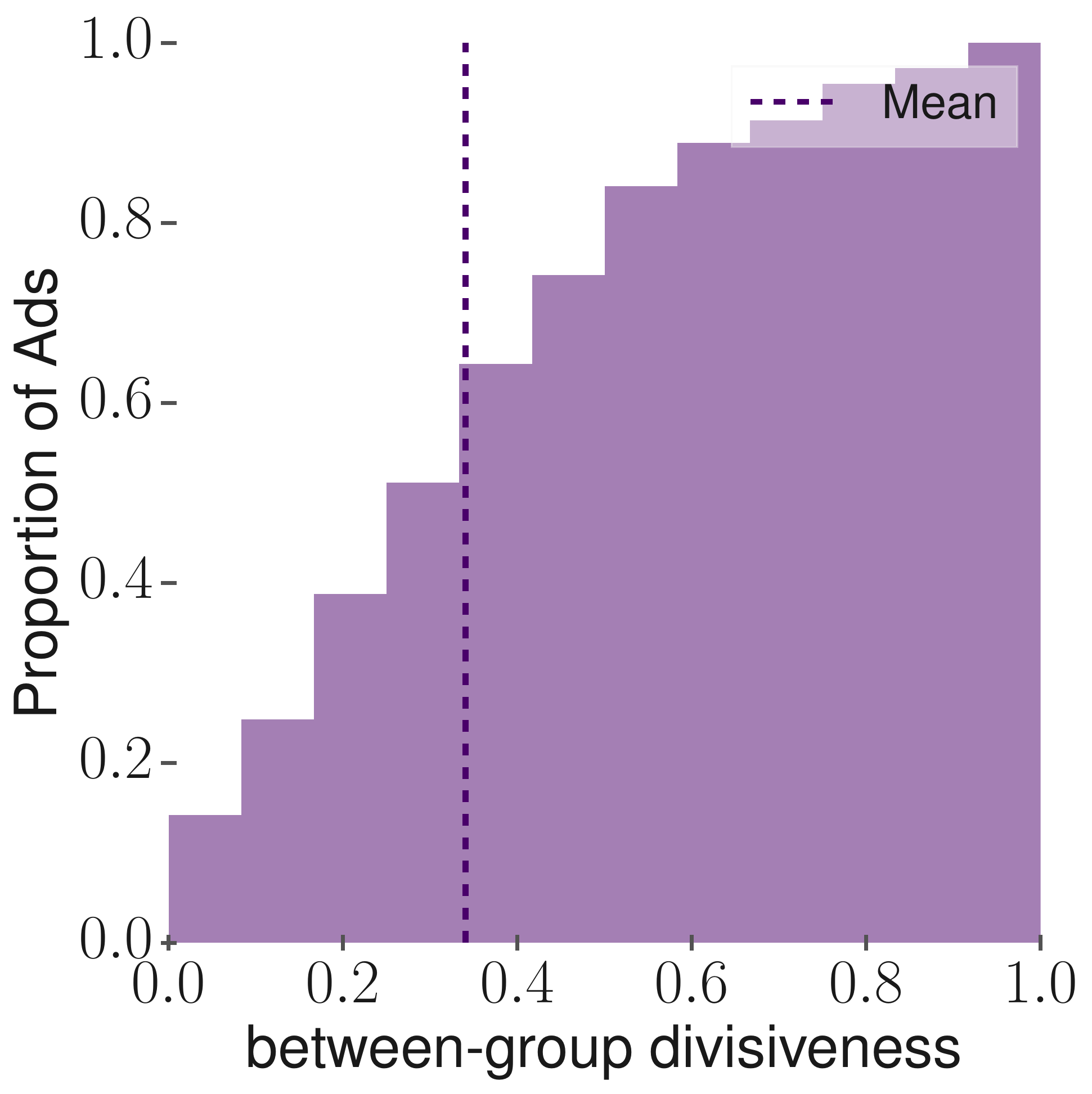}
    \caption{Between group divisiveness}
    \end{subfigure}\hfill
\begin{subfigure}[b]{0.50\columnwidth}
    \centering
  \includegraphics[width=\columnwidth]{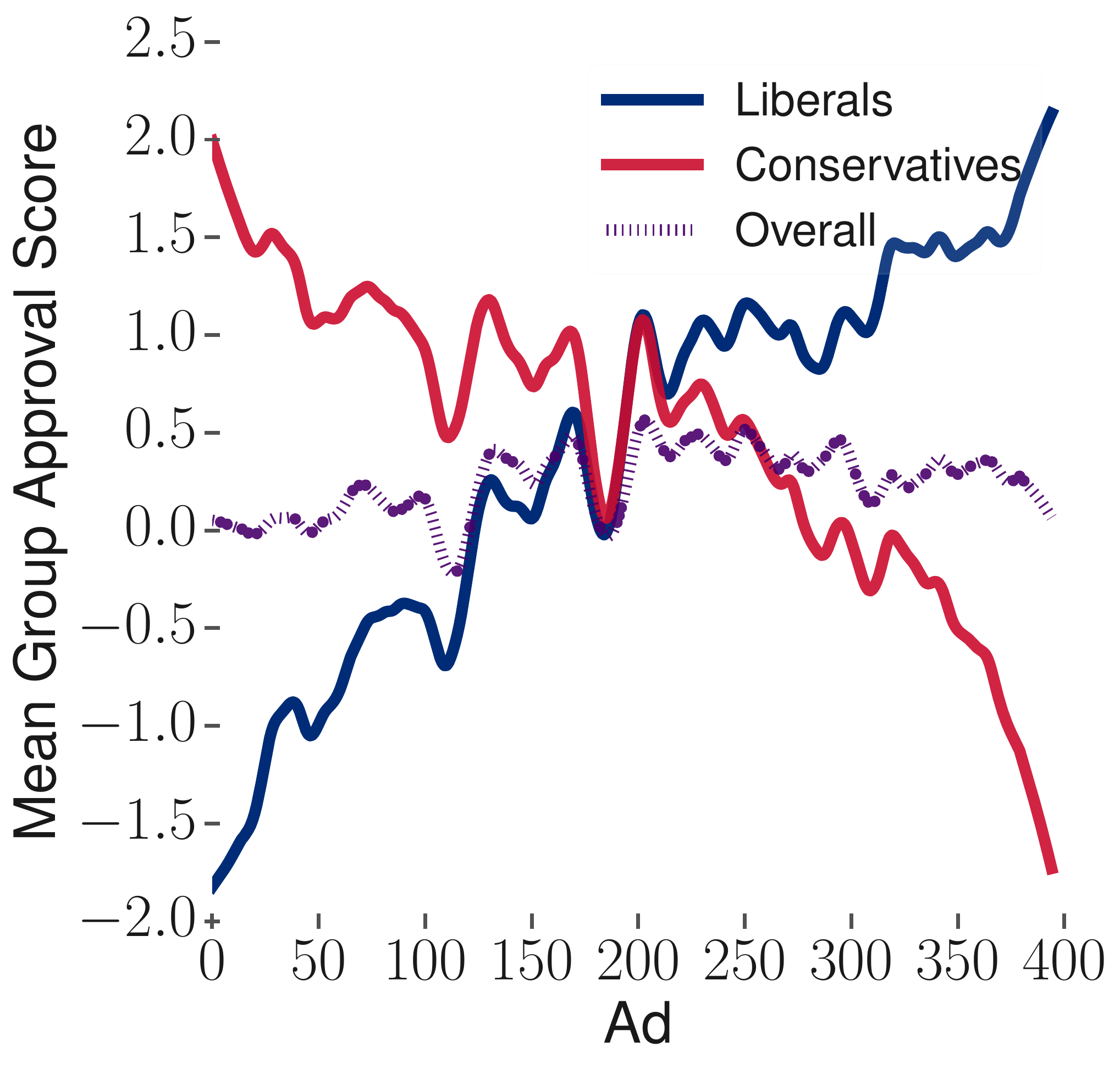}
    \caption{Ideological Differences}
    \end{subfigure}\hfill
    \caption{Distribution of the ads on approval and disapproval: (a\&b) overall, (c\&d) across ideological groups. (a\&c) plot the cumulative distribution functions (cdfs), (b\&d) plot the differences in approval in each ad, where x-axis consists of all the ads.}
    \label{fig:approvalContent}
\end{figure}

As another characterization of people's reactions to the ads, we asked respondents in a second survey whether they approve or disapprove of a particular ad, and how strongly they approve or disapprove.\footnote{Specifically, we asked ``Do you approve or disapprove of what the ad says or implies?'' Answer choices: Approve; Disapprove; Neither; There is nothing in this ad to approve or disapprove of; I don't know. Followed by a measure of strength ``Do you [approve/disapprove] very strongly, or not so strongly?'' if the prior question was answered with approve or disapprove.} These questions in the survey were constructed based on questions about political preference that have been extensively pre-tested by Pew Research for previous surveys about political polarization~\cite{beatty2007research}. 
We find that 87\% of the adds were approved and 63\% of the ads were disapproved by at least 20\% respondents (see Figure~\ref{fig:approvalContent} (a)).
To quantify the received responses, we assigned an approval score on a 5 point scale with values of -2 (strong disapproval), -1 (weak disapproval), 0 (neither approve or disapprove), +1 (weak approval), and +2 (strong approval). While computing the mean approval score for a group, we dropped the 0 responses to ensure that a mean approval score close to 0 corresponds to similar weights from approval and disapproval. Table~\ref{table:approvalExamples} lists some example ads along with their approval tendencies by the two ideological groups within our dataset.

Figures~\ref{fig:approvalContent} (c\&d) show the relationship between respondents' ideology and approval of ad content. We observe that the mean within-group divisiveness for liberals is 0.92 (stdev = 0.36) and 0.98 (stdev = 0.31) for conservatives (Table~\ref{table:divisivenessSummary}). Both the within-group divisiveness values being lower than 1, suggests that the likelihood with which individuals within the same ideological group would agree about approving an ad is higher than that when compared against individuals across ideological groups.
The divisiveness in approval responses is further confirmed by the between-group divisiveness measure which ranges between 0 and 1 (mean = 0.34) across the high impact ads.

\begin{table}[t!]
\renewcommand{\arraystretch}{1.1}
\centering
\tiny
	\begin{tabular}{@{}p{\columnwidth}@{}}
\toprule
Approved by both the liberals and the conservatives\\
\hline
\textit{Show up, fight racism and take a stand for equality. Monday, May 2 at 4 PM at Erie County Holding Center Justice For India: Not 1 More!}\\
\rowcollight \textit{Alton Sterling, an innocent 37-year-old Black male, was outrageously executed by two Baton Justice For Alton Sterling}\\
\textit{Did you see this? Damn... We lost count of how many mentally ill citizens were murdered during encounters with violent cops. Here is another woman suffering from mental illness. Full story: http://bit.ly/10rglhk}\\
\rowcollight \textit{Join Us! Support The Police!}\\
\textit{Darkness cannot drive out darkness; only light can do that. Hate cannot drive out hate; only love can do that. Martin Luther King, Jr.}\\

\hline
Disapproved by both the liberals and the conservatives\\
\hline
\textit{No wonder white boys don't get shot when they're arrested':}\\
\rowcollight \textit{Anti-immigration is the only salvation!}\\
\textit{O000ps seems like someone screwed up! Salute our brave and smart cops who mistake man's member for a deadly weapon. Follow US and stay WOKE!}\\
\rowcollight \textit{It's ok they're women so they'll only find the kitchen}\\
\textit{This man beat up police officer who tased his wife. Do you agree with the man who defended his woman?}\\

\hline
Approved by the liberals and disapproved by the conservatives\\
\hline
    \textit{Two years have passed since August 11, 2014, the date. when 25-year-old Ezell Ford was mur Justice For Ezell Ford And Donnell Thompson}\\
\rowcollight \textit{We don't want to honor racism, slavery and hatred. This is what Confederate Heritage is. Not My Heritage Rally}\\
\textit{Say it loud: I'm black and I'm proud!}\\
\rowcollight \textit{We Muslims of the United States are subject to Islamophobia from the media where regularly STOP SCAPEGOATING MUSLIMS!}\\
\textit{Click to Learn More! Everybody knows that Islam is against terrorism but not everyone believes this fact! Islam does not support terrorism under any circumstances. Terrorism goes against every principle in Islam. In fact if a Muslim engages in terrorism, he is not following Islam and so he is not a Muslim! America, stop insulting peaceful citizens, stop taking all of us as criminals, we don't deserve such attitude! \#muslimvoice \#muslim @muslim Voice Musliminst}\\
\hline
Approved by the conservatives and disapproved by the liberals\\
\hline
\textit{Heritage not hate y'all! Our flag has nothing to do with racism! The Federal Government shouldn't be able to dictate what we can and cannot do. Go follow Confederate page \#1 on Instagram south united if you are proud of our southern heritage. God bless Dixie! Confederate page \#1 on Instagram! No racism, no hate! The south will rise again!}\\
\rowcollight \textit{If we ever forget that we are One Nation Under God, then we will be a nation gone under. Ronald Reagan}\\
\textit{Our country was drawing a blank for the last eight years. We need a strong leader who will March for Trump}\\
\rowcollight \textit{America is at risk. To protect our country we need to secure the border.}\\
\textit{Stop refugees! The're taking our jobs!}\\
    \bottomrule
\end{tabular}
    \caption{Example ads on the basis of the approval behavior by the respondents from two political ideologies.}
\label{table:approvalExamples}
\end{table}


\subsection{Perceptions of false claims in the ads}

\begin{figure}[t!]
\centering
    \begin{subfigure}[b]{0.50\columnwidth}
    \centering
  \includegraphics[width=\columnwidth]{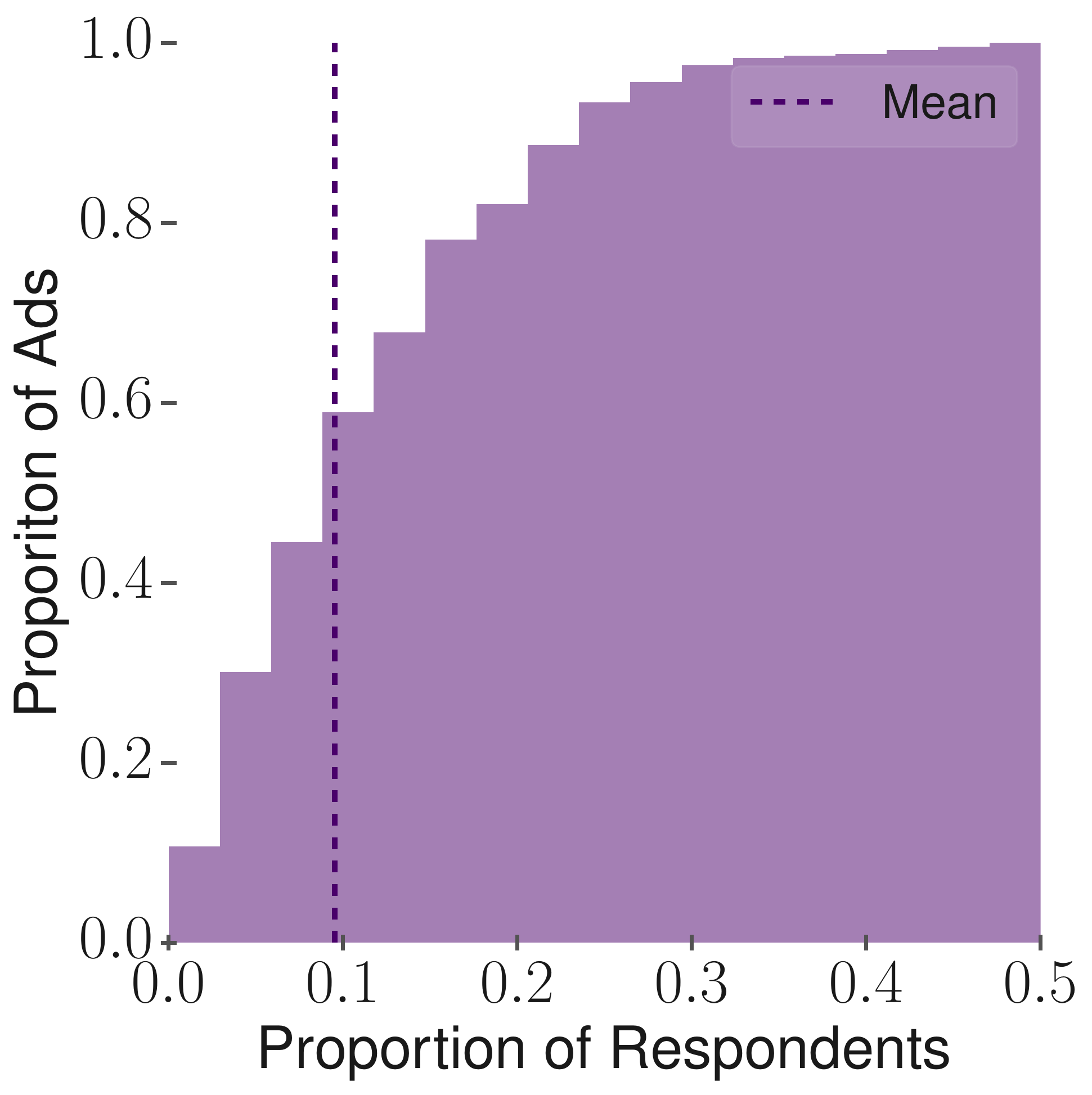}
    \caption{Proportion of FCs Identified}
    \end{subfigure}\hfill
    \begin{subfigure}[b]{0.50\columnwidth}
    \centering
  \includegraphics[width=\columnwidth]{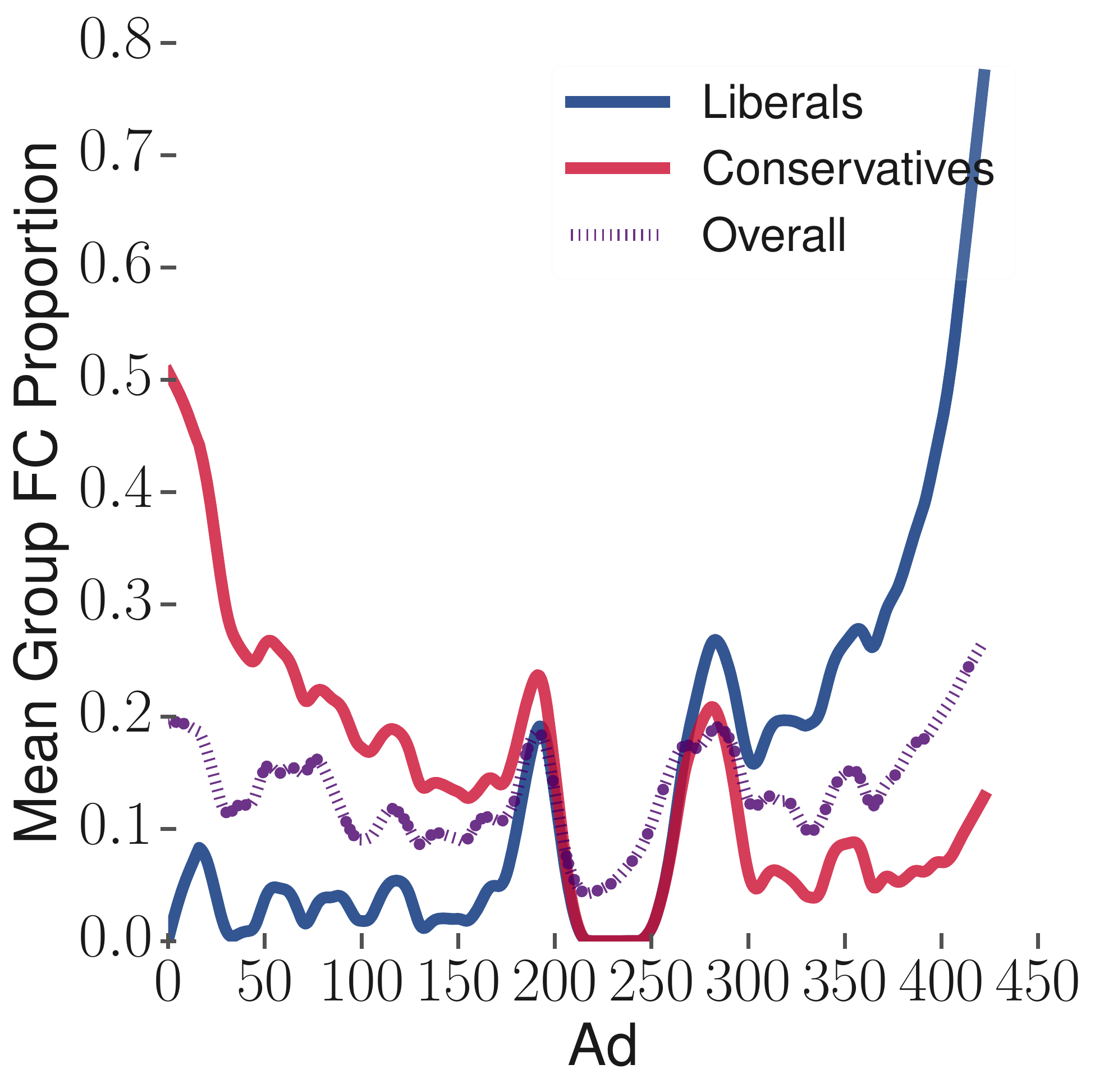}
    \caption{Political Ideology}
    \end{subfigure}\hfill
    \caption{Distribution of the ads on false claims (FCs): (a) overall (as a cumulative density function), (b) across ideological groups (where each ad is plotted on the $x$-axis).}
    \label{fig:fcContent}
\end{figure}

To examine whether the high impact IRA ads contained any false claims, in another survey we asked the respondents if they could identify any false claims present in the ads.\footnote{Specifically, we asked respondents to ``Please copy and paste any phrases or sentences in the advertisement that you think contain a ``factual claim''. That is, something that someone could verify as True or False. If you cannot identify any claims, please type ``No Claims'' in the first box.'' We then asked them to label the phrases they had identified as ``True'', ``False'' or ``not sure whether they are True or False''.} We find that 89\% (433 out of 485) of the high impact ads were identified to have at least one false claim, and about 45\% of the ads contained false claims according to 10\% of the respondents. Figure~\ref{fig:fcContent} (a) shows the cumulative distribution of the ads with the number of respondents who identified at least one false claim in them.

Next, as in the other two content analyses, we examined whether respondents' ideology related to their perception of the presence of false claims (Figure~\ref{fig:fcContent}). Both the within-group divisiveness values being lower than 1, suggests that the likelihood with which individuals within the same ideological group would agree about finding false claim in an ad is higher than that when compared against individuals across ideological groups.
Table~\ref{table:fcExamples} shows a sample of ads and false claims identified by respondents from each ideological group (liberal, conservative). 

\begin{table}[t]
\renewcommand{\arraystretch}{1.1}
\centering
\tiny
	\begin{tabular}{p{\columnwidth}}
\toprule
False Claims identified by both  liberals and conservatives\\
\toprule
\textit{\hlfc{Bernie Sanders has proven himself to be the best candidate in every way}. He is fair and strong and he is the \hlfc{only one fighting for the black community}. For more than 40 years he is advocating against any injustice and stays true to his moral values. Vote for Bernie!}\\
\rowcollight \textit{\hlfc{Illegal immigrants are not only flooding our country with drugs and sinking our economy, they are a major threat to Americas integrity. They don't care about American laws, history and heritage. They just want our money and social benefits.} Considering the fact that they multiply like rabbits, soon we will all wear sombreros. We need to stop this invasion.}\\
\textit{When you live in Texas \hlfc{you know that you are the chosen one!}}\\
\rowcollight \textit{It is time to wake up and see the truth. \hlfc{Cops are not our friends and government doesn't care about you}. Share this if you're awake!.}\\
\textit{\hlfc{It's ok they're women so they'll only find the kitchen}}\\
    
\hline
False claims identified by  liberals.\\
\hline
    \textit{Clinton said the United States needs to confront the 'systematic racism' in its law enforcement efforts. We're sick of politicians organizing and leading the systematical propaganda against our police. It is unfair and vilely to accuse our heroes of every sin and crime. In fact, the \hlfc{efficiency of our cops resulted in a decrease of the average amount of crimes}, especially in large cities. Law-abiding citizens should never fear cops, but criminals do. And that's why \hlfc{Hillary is on the criminals' side.} Join our rally on July, 23th in New York City, it's time to show Clinton that we will never let her become our next President!}\\
\rowcollight \textit{It might sound like a cliche but "get a job" is a really good advice for young liberals protesting against everything in the world. Old man Ronald knew what he was talking about! Our college students should have an experience of paying taxes before standing for illegal immigrants' rights. \hlfc{They should rise their own children before standing for gay parenthood}. \hlfc{It's no secret most active liberal's supporters are people about 20-25 years- old while most conservatives are older}. Well; as they say; wisdom comes with ages.}\\
\textit{His \hlfc{failed medical reform and unbelievable national debt is enough to put Obama behind bars. but that's not all}. His greatest "accomplishment" is flooding America with countless criminals and giving them all an absolute omnipotence. \hlfc{Thanks to Barack Hussein Obama we have at least one big terror attack each year; not to mention illegals raging out and poisoning our country with drugs}. For what he did to America Obama should rot in prison for the rest of his life.}\\
\rowcollight\textit{Border Patrol agents in South Texas arrested an illegal alien from Honduras that had previously been deported and convicted of Rape Second Degree. \hlfc{Thanks to Obama's and Hillary's policy, illegals come here because they wait for amnesty promised}. The wrong course had been chosen by the American government; but all those \hlfc{politicians are too far from the border to see who actually sneaks through it illegally}. Rapists, drug dealers, human traffickers; and others. The percent of innocent poor families searching for a better life is too small to become an argument for amnesty and Texas warm welcome.}\\
\textit{\hlfc{Anti-immigration is the only salvation!}}\\
\hline
False claims identified by conservatives.\\
\hline
\textit{Don't Shoot is a community site where you can find recent videos about \hlfc{outrageous police misconduct}, really valuable ones but \hlfc{underrepresented by mass media}. We provide you with first-hand stories and diverse videos. Join us! Click Learn more!}\\
\rowcollight \textit{We don't want to honor racism, slavery and hatred. \hlfc{This is what Confederate Heritage is}. Not My Heritage Rally}\\
\textit{The USA is exactly the place \hlfc{where cops can't care less about people's civil rights. They are cynical toward the rule of law and disrespectful of the rights of fellow citizens}. Details: http://donotshoot.us/}\\
\rowcollight \textit{\hlfc{Police are beyond out of control}, help us make this viral! Follow our account in order to spread the truth!}\\
\textit{Join us to study your blackness and get the power from your roots. \hlfc{Stay woke and natural! Nefertiti's Community}}\\
    \bottomrule
\end{tabular}
    \caption{Example ads on the basis of false claims identified by the respondents from two political ideologies. Identified false claims are highlighted in pink. }
\label{table:fcExamples}
\end{table}


\subsection{Summary}
This section focuses on peoples' perceptions of the content of the 485 IRA ads we identified as high impact. To assess these perceptions along three axes -- likelihood of being reported, approval and disapproval, and the presence of false claims -- we conducted three U.S. census-representative surveys. Our analysis of the perceptions queried in these surveys shows that ideological opinions of individuals influence their perceptions of these ads. We find that many of these ads were severely divisive, and generated strongly varied opinions across the two ideological groups of liberals and conservatives (see Figure~\ref{fig:reportingContent},~\ref{fig:approvalContent},~\ref{fig:fcContent}).

\section{Analyzing the Targeting Formula}

Next, we focus on understanding how the target formula is created by advertisers and the role that Facebook interface plays on that. 


\subsection{Targeting Possibilities}


The Facebook ads platform provides three approaches for advertisers to target people~\cite{speicher-2018-targeted,Andreou18a}, briefly described next.

\vspace{1mm}
\noindent \textit{Personally Identifiable Information (PII)} targeting is the form in which advertisers provide personal information about users such as name, phone number, and email address so that Facebook can directly place the ads to them. This kind of targeting does not appear in the IRA dataset.

\vspace{1mm}
\noindent \textit{Look-alike audience target}. For this targeting option, advertisers provide to Facebook a list of users similar to that one in the PII or a list of people who liked the advertiser Facebook page. Then, Facebook 
attempt to target a similar audience to the group in this specific list. Only $1.1\%$ of the \textit{high impact} ads used this option.

\vspace{1mm}
\noindent \textit{Attribute-based targeting} allows the advertiser to create a target formula based on a wide range of elements that include user basic demographics (i.e. gender, age, location, language), advanced demographics (i.e. political leaning, income level, `Parents with children preschoolers'), interests (i.e. newspapers, religion, politics), and behaviors (i.e. `Business Travelers' or `New Vehicle buyers').  Recent work showed that the number of possible interests provide by Facebook is greater than 240,000~\cite{speicher-2018-targeted}. Facebook allows one to include or exclude users with each of those attributes and combine multiple attributes as part of a target formula. The vast majority of the \textit{high impact} ads, $895$ out of the $905$, used this option to elaborate a formula. 
We found that $78\%$ of the ads used $2$ or more interests and behaviors in their formula, creating very complex formulas with up to $39$ distinct attributes.

\begin{figure}[!t]
    \centering
    \includegraphics[width=1\linewidth]{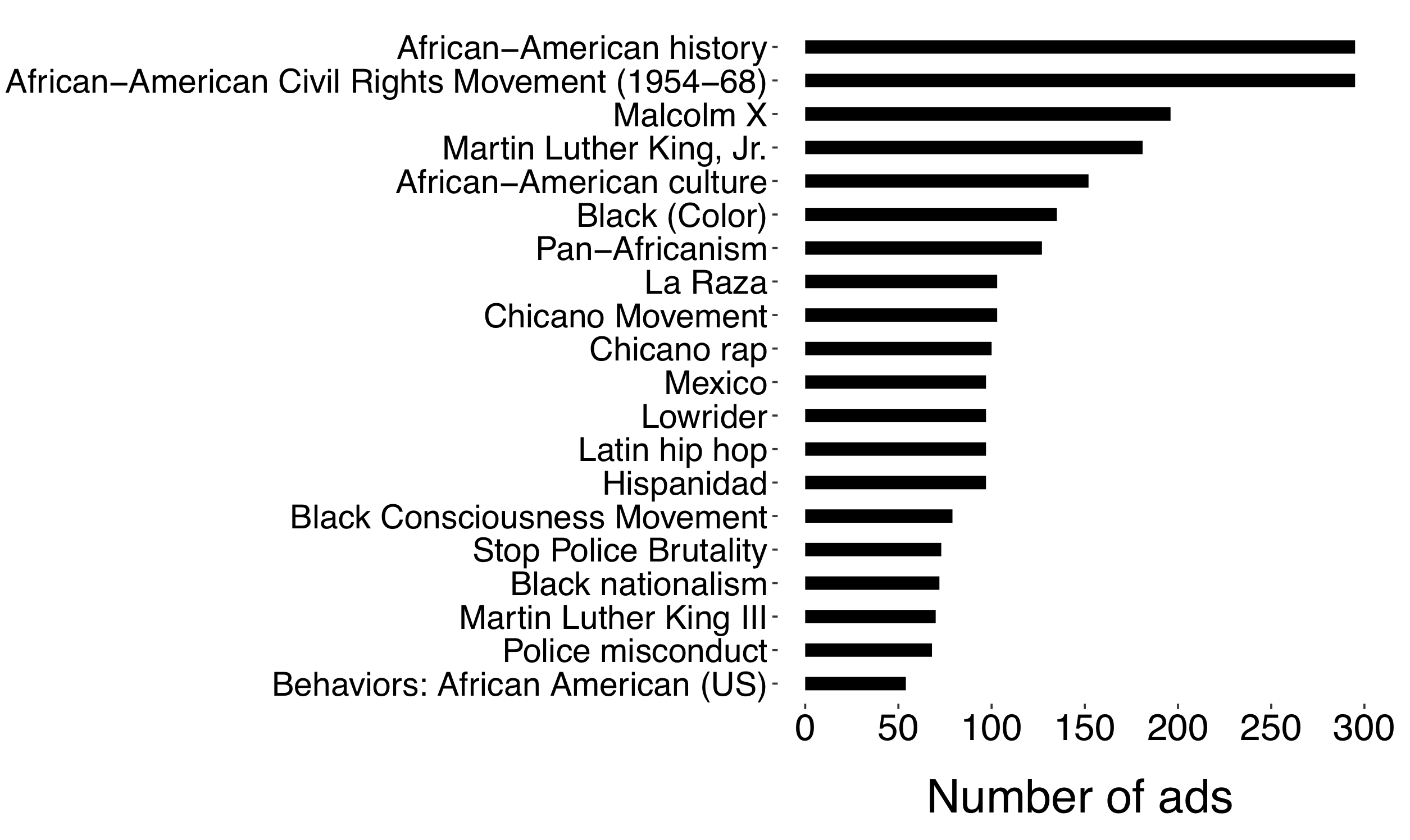}
    \caption{Top 20 attributes based on the number of advertisements they appeared.}
    \label{fig:topInterests}
\end{figure}

Figure~\ref{fig:topInterests} shows the top attributes that appear in the ads target formula based on the number of times they appeared in different ads. There were 497 distinct attributes and the most present attributes interest were African-American history and African-American Civil Rights Movement (1954-68), appearing in 295 (32\%) ads. We can note a prevalence of attributes related to African-American and Hispanic Population, with interests like Mexico, `Hispanidad' and `Latin hip hop'. Next, we investigate aspects of the Facebook ads platform design that might have favored the IRA ads to massively explore this particular targeting strategy.

\subsection{The Role of Attribute Suggestions}

Facebook provides a tool for advertisers that, given a target attribute, it presents a list of other attributes that target people with similar demographic aspects~\cite{speicher-2018-targeted}.  For example, in the list of suggested targeting interests for `Townhall.com', a page with an audience in which $79.5\%$ of the users are very conservative users according to Facebook, there are other pages with similar bias towards very conservative users, i.e. `The Daily Caller' ($67.1\%$), `RedState' ($84.3\%$), and `TheBlaze' ($59.6\%$)~\cite{ribeiro2018@icwsm}. 

\begin{figure}[!t]
     \centering
     \includegraphics[width=0.7\columnwidth]{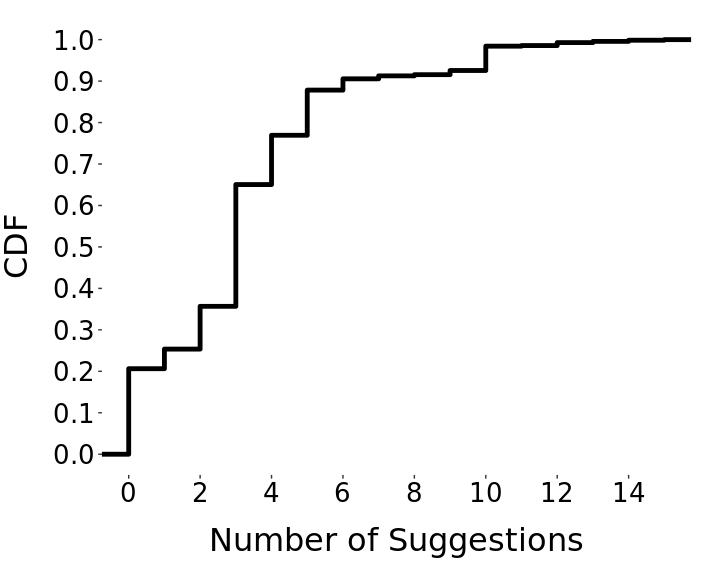}     
     \caption{\textbf{Cumulative Distribution Function (CDF) for the number of suggestions.}}
     \label{fig:suggestions}
 \end{figure}

In order to investigate if the IRA ads have used suggestions to elaborate complex targeting formulas, we crawled the attribute suggestions for each attribute that appear in the dataset of highly impact ads. Figure~\ref{fig:suggestions} shows the cumulative distribution function for the number of suggested attributes that appear in the same formula.  We can see that around 64\% of the ads that potentially used this feature because they have at least three target attributes suggested by Facebook as part of the same formula. There are 1.2\% of ads with more than 10 suggested attributes in the same formula. 
As an example, all the 13 interests, including Islam, Ramadan, Islamism, used in the target formula of the ad ID 1915\footnote{http://www.socially-divisive-ads.dcc.ufmg.br/app.php?query=1915} appear as suggestions for at least one of the others in the formula. For ad ID 1840\footnote{http://www.socially-divisive-ads.dcc.ufmg.br/app.php?query=1840}, we were able to find 9 out of 10 of the interests using the interest suggestion feature. This provides evidence that this feature may have been a key element used by the IRA campaign to choose the target audience. 

\subsection{Summary}

In this Section we show that the vast majority of the IRA ads use attribute-based targeting, containing complex target formula that includes interest and behavioral attributes that are likely suggested by Facebook. Next, we investigate the extent to which these formulas allowed advertisers to reach demographic biased audiences.

\section{Analyzing the Target Audience}




We start by describing our methodology to reproduce the IRA queries (without running the ad) and gather the demographics of the of the targeted users. 



\subsection{Assessing the Audience Demographics}

Before launching an advertisement in Facebook, the advertiser can get the estimated audience (i.e., the number of monthly active users) likely to match the target formula. Our methodology consists of using the Facebook Marketing API\footnote{developers.facebook.com/docs/marketing-apis} to reproduce the targeting formula of all high impact IRA ads and get the demographics of the population that matches each targeting formula, without running any ad. This methodology has been extensively used recently for different purposes, including inferring news outlets political leaning~\cite{ribeiro2018@icwsm}, study migration~\cite{zagheni2017leveraging} and gender bias~\cite{garcia2017facebook} across countries, and for public health awareness~\cite{saha2017characterizing} and lifestyle disease surveillance~\cite{araujo2017using}.  For our analysis, we considered seven demographic categories: political leaning, race, gender, education level, income, location (in terms of states), and age. As a baseline for comparison, we also gathered the demographic distribution of the United States Facebook population.

Only 11\% of the used attributes that appear in the IRA ads targeting formulas are not available for targeting anymore due to changes in the Facebook Marketing API. 
In most of these cases, we reproduced the ad target formula without the missing attribute, especially when the attribute looks redundant with the others in the formula. 
We did not reproduce only 6 targeting formulas.

\subsection{Measuring Audience Bias}

To assess the audience bias of each of the demographic aspects that we considered, we computed the differences between the fraction of the population with a demographic aspect and the same fraction of the population in the baseline distribution (i.e. the U.S. Facebook population), namely the \textit{bias score}. For instance, if the percentage of African-Americans in the audience of a particular ad is 40\%, the \textit{bias score} for this dimension in the ad is 0.25 as the percentage of African-American in the U.S. Facebook population is nearly 15.5\% ($0.4-0.155$). 


\begin{figure}[t]
    \centering
    \includegraphics[width=1\linewidth]{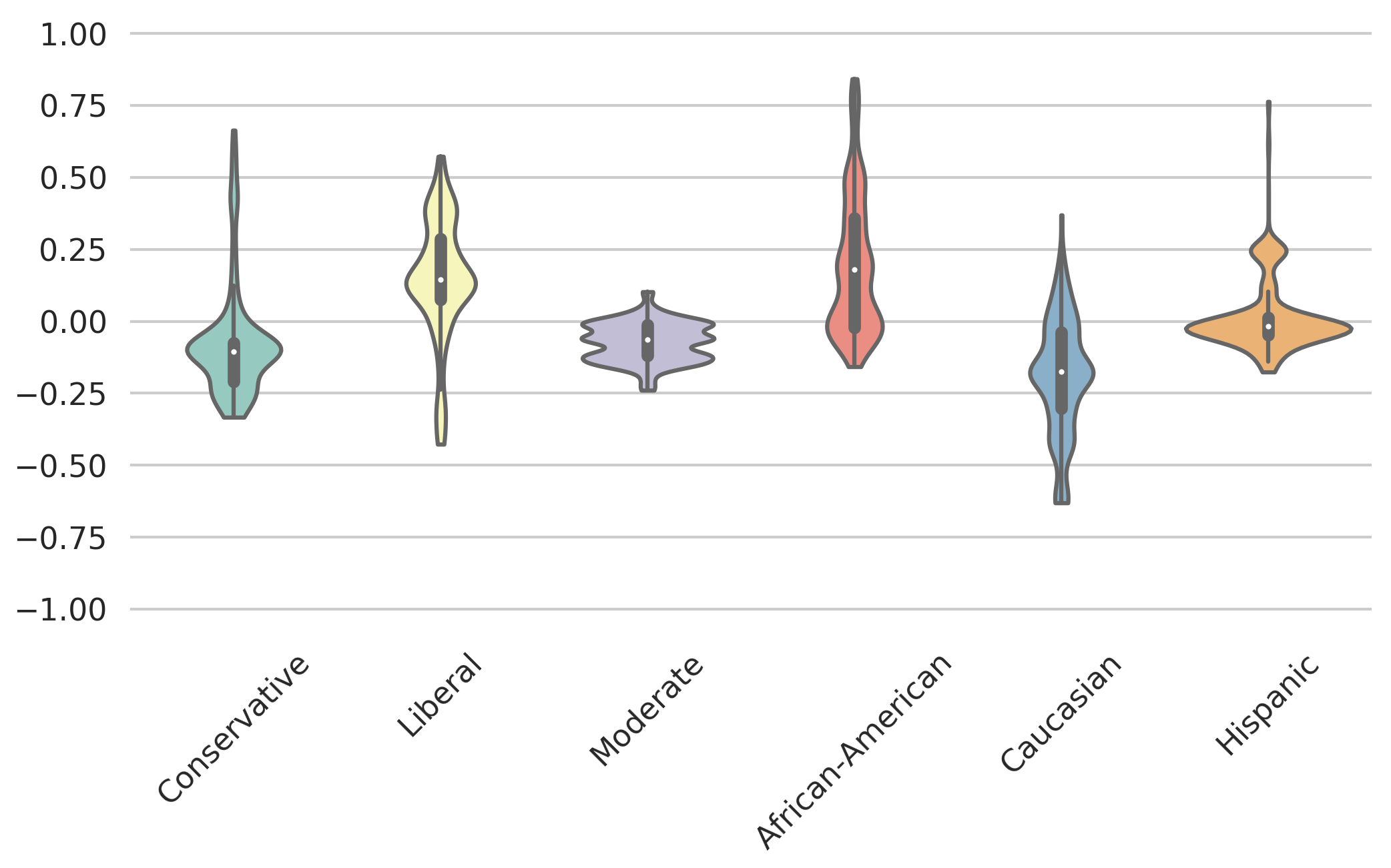}
    \caption{\textbf{Bias in demographic dimensions. Each violin represents the bias score for all high impact ads in a particular demographic dimension. The median is represented by a white dot in the center line of the violin. 50\% of the data is present between the two thick lines around the center}}
    \label{fig:violin}
\end{figure}

Figure \ref{fig:violin} depicts the distribution of the measured bias on political leaning and ethnic affinity.
In comparison with all the demographic categories, these two showed to be the ones with the highest biases. We note that most of the ads target audiences that are more biased towards the African-Americans population and the Liberals. More specifically, about 70\% of the IRA ads target an audience with a higher proportion of African-Americans than in the US Facebook distribution. This difference is even accentuated for Liberals, with 82\% more biased in comparison with the reference distribution. 
The percentage of ads with bias score superior to 0.15 is 52\% for African-American and 41\% for Liberals.
Our dataset suggests the presence of those ads that target extremely biased populations of conservatives, Liberals, Hispanic, and especially African-Americans. The target audiences for the IRA ads are slightly biased towards women and young adults (18-34 years), which are omitted from Figure~\ref{fig:violin} due to space constraints.

\subsection{Targeting audience and Divisiveness}


Next, we investigate if the
advertisers target the ads towards audiences that are less likely to identify their inappropriateness due to their ideological perception bias.
Additionally, we examine if the ads directed to biased audiences could leverage the already existing societal divisiveness to further amplify it among the masses.

\begin{table}[t!]
\centering
\small
    \begin{tabular}{lrrr}
    \toprule
 Group & Report & Approval & False Claims\\
    \toprule
    Liberals & -0.17*** &  0.41*** & - \\
    Conservatives & -0.15*** & 0.32***&  - \\
\bottomrule
\end{tabular}
\caption{Pearson's $r$ correlation between targeting and the ideological divisiveness for the high impact ads (*** $p<0.001$, no statistical significance in the case of false claims).}
\label{table:correlationTargetingSummary}
\vspace{-0.2in}
\end{table}


\begin{figure}[t]
    \centering
    \begin{subfigure}[b]{0.5\columnwidth}
    \centering
  \includegraphics[width=\columnwidth]{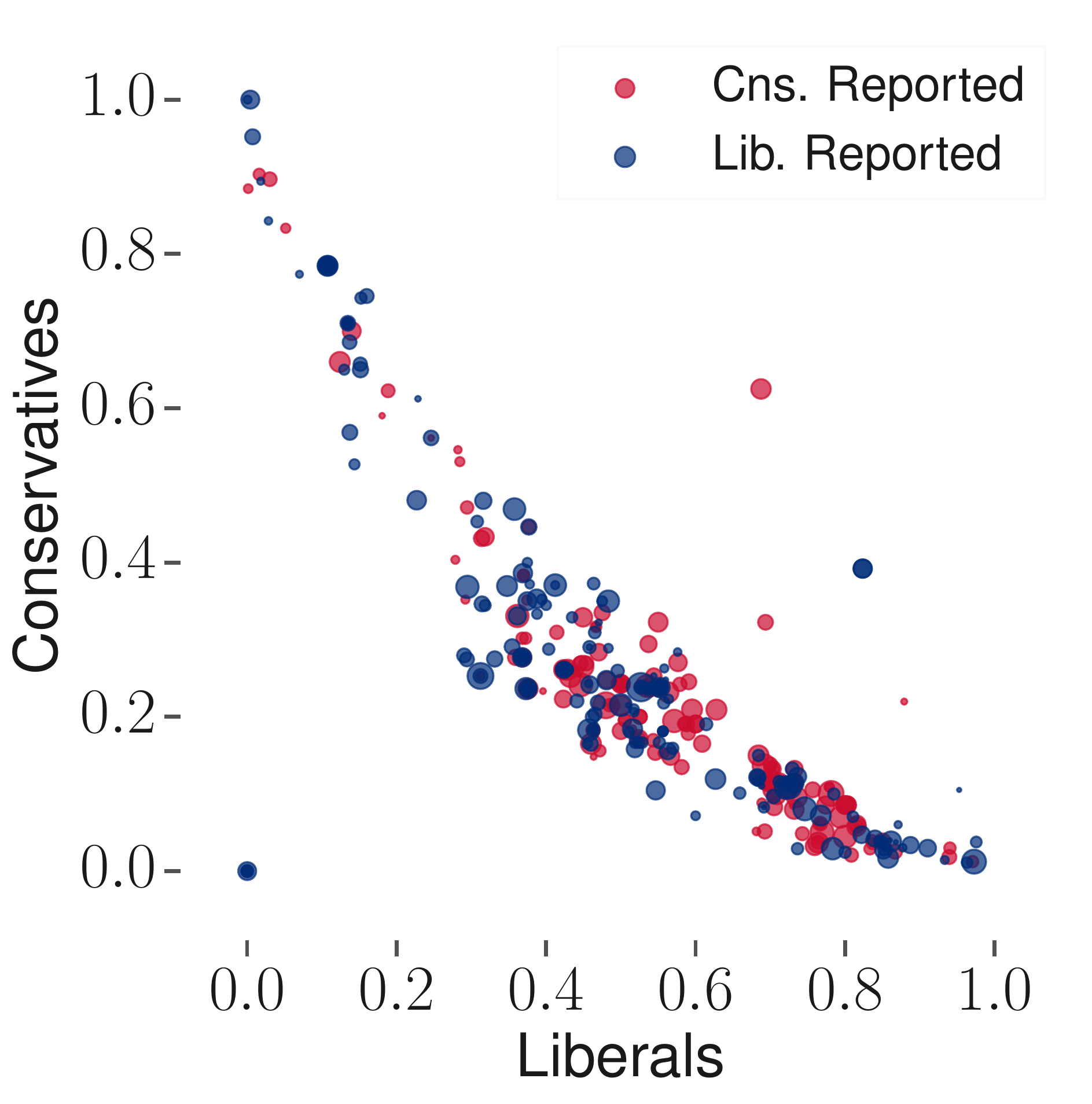}
    \caption{Reporting: Targeted}
    \end{subfigure}\hfill
        \begin{subfigure}[b]{0.5\columnwidth}
    \centering
  \includegraphics[width=\columnwidth]{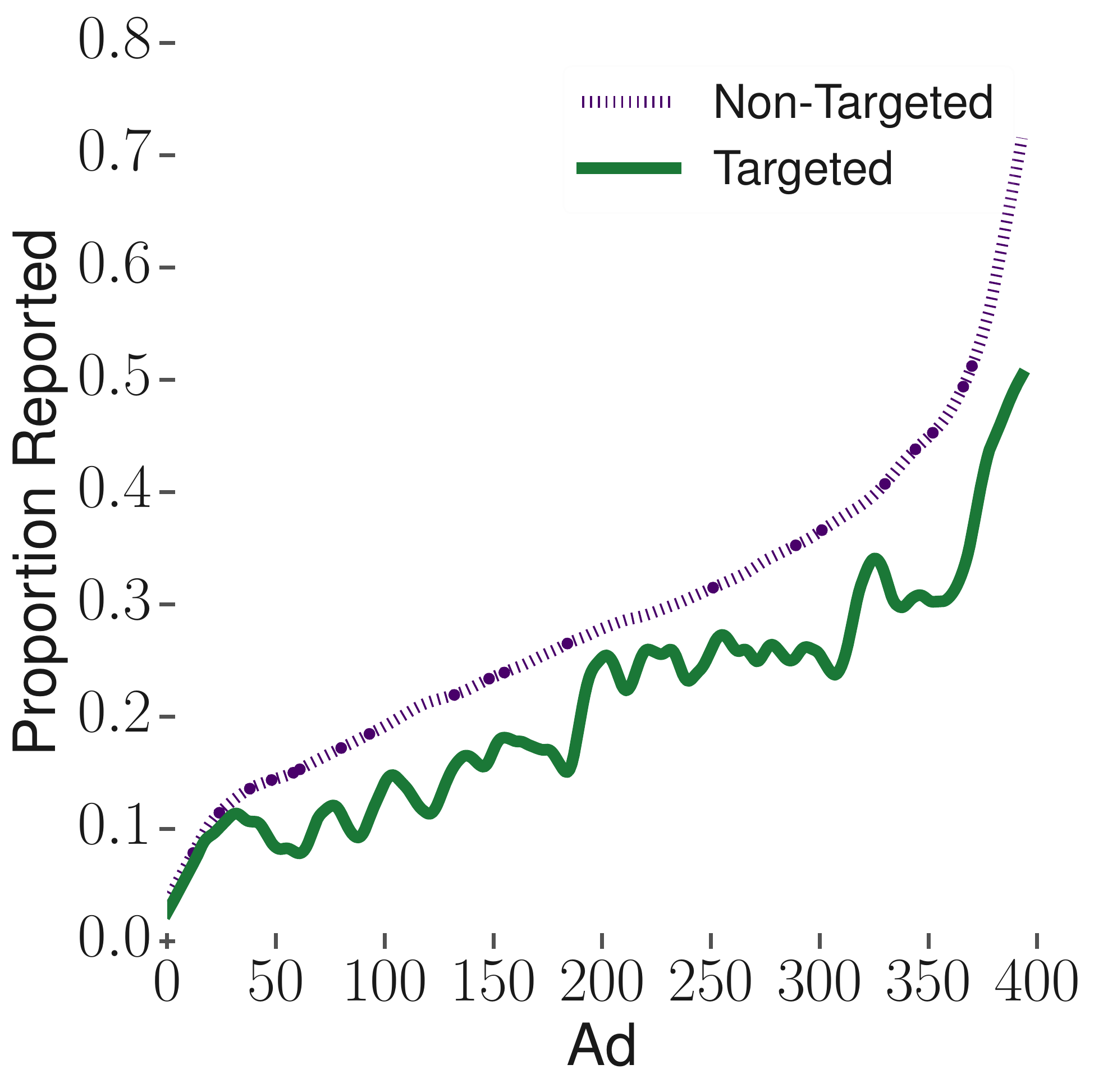}
    \caption{Reporting: Bias}
    \end{subfigure}\hfill
    \begin{subfigure}[b]{0.5\columnwidth}
    \centering
  \includegraphics[width=\columnwidth]{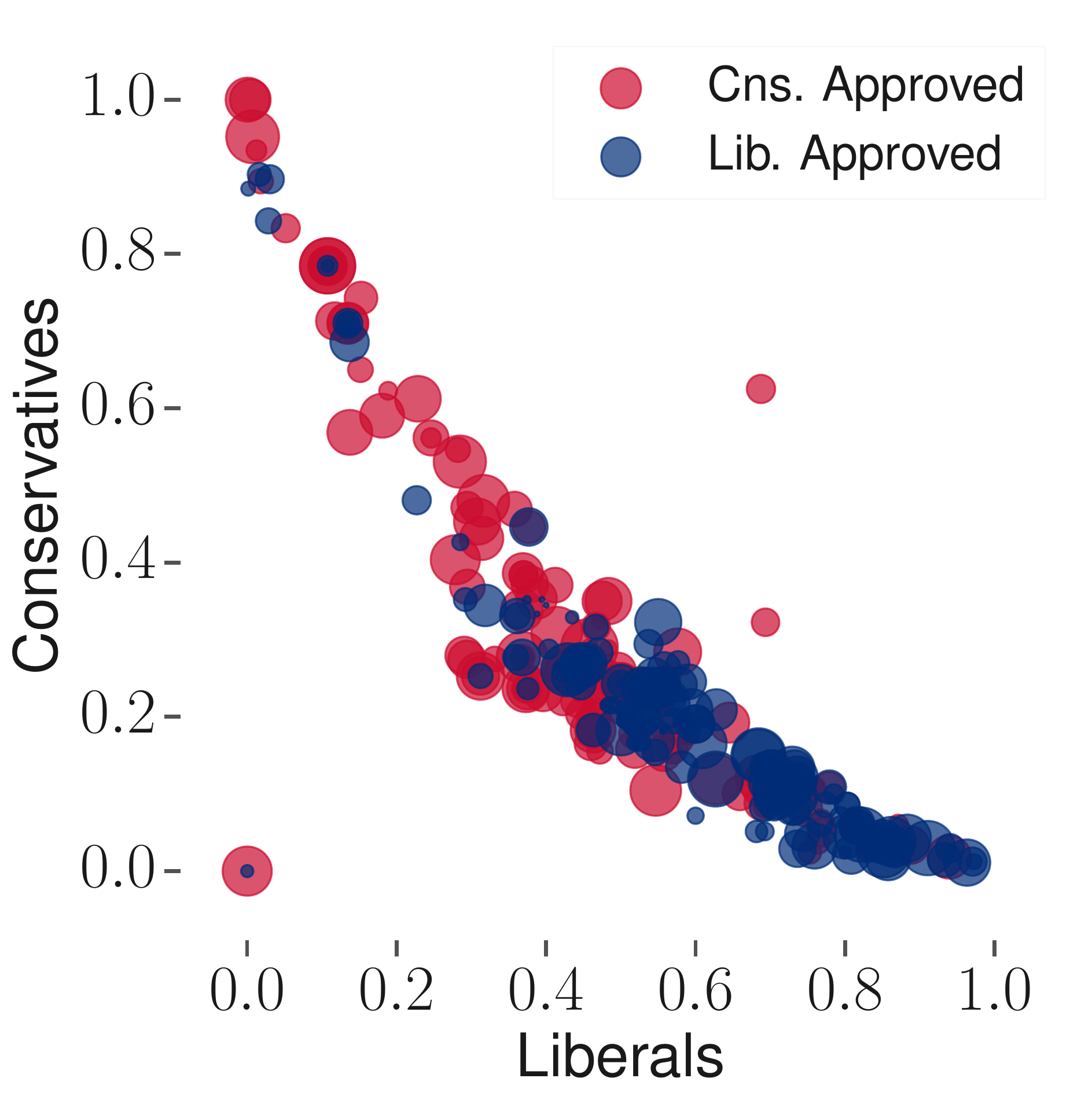}
    \caption{Approval: Targeted}
    \end{subfigure}\hfill
        \begin{subfigure}[b]{0.5\columnwidth}
    \centering
  \includegraphics[width=\columnwidth]{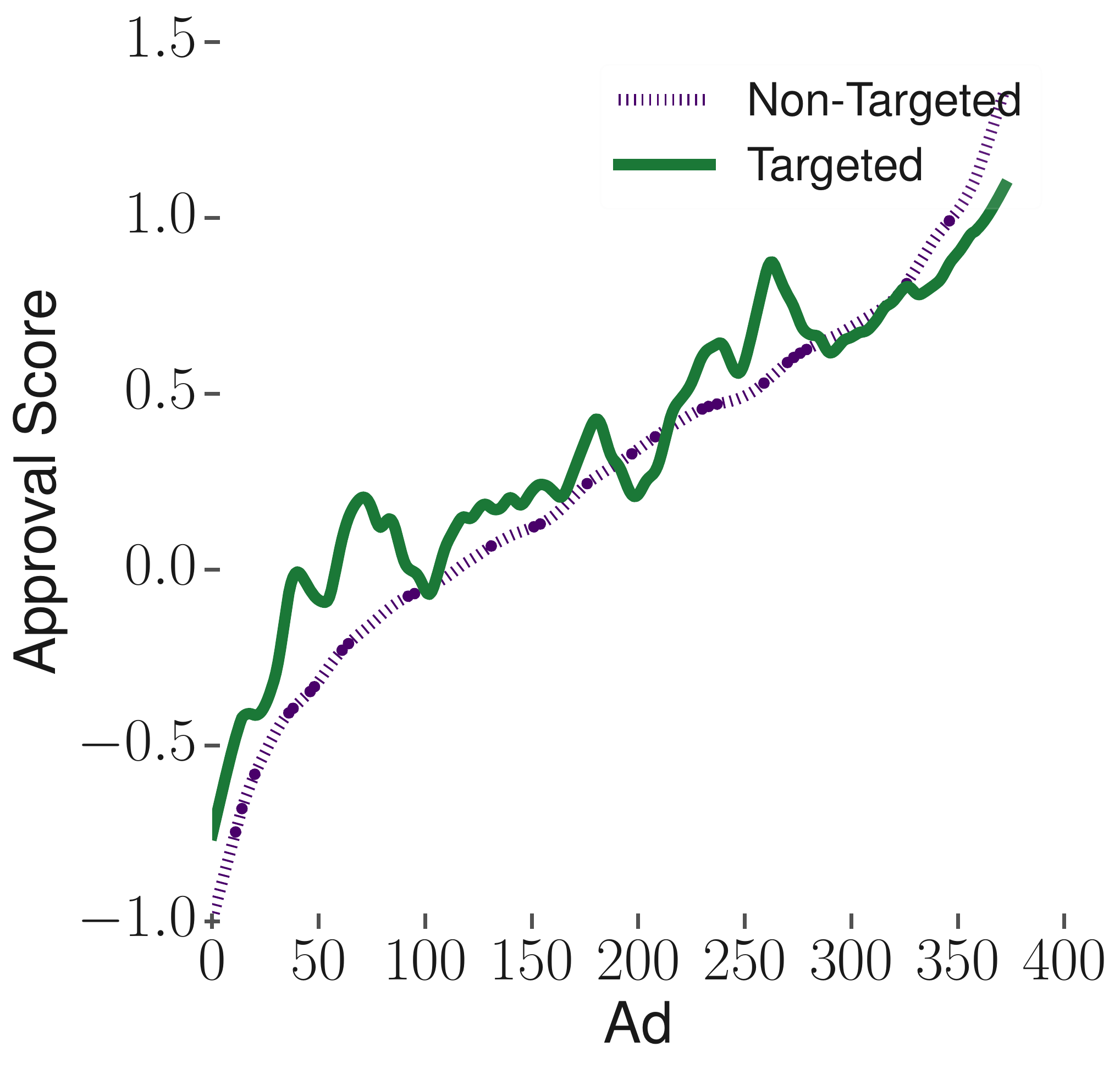}
    \caption{Approval: Bias}
    \end{subfigure}\hfill
\begin{subfigure}[b]{0.5\columnwidth}
    \centering
  \includegraphics[width=\columnwidth]{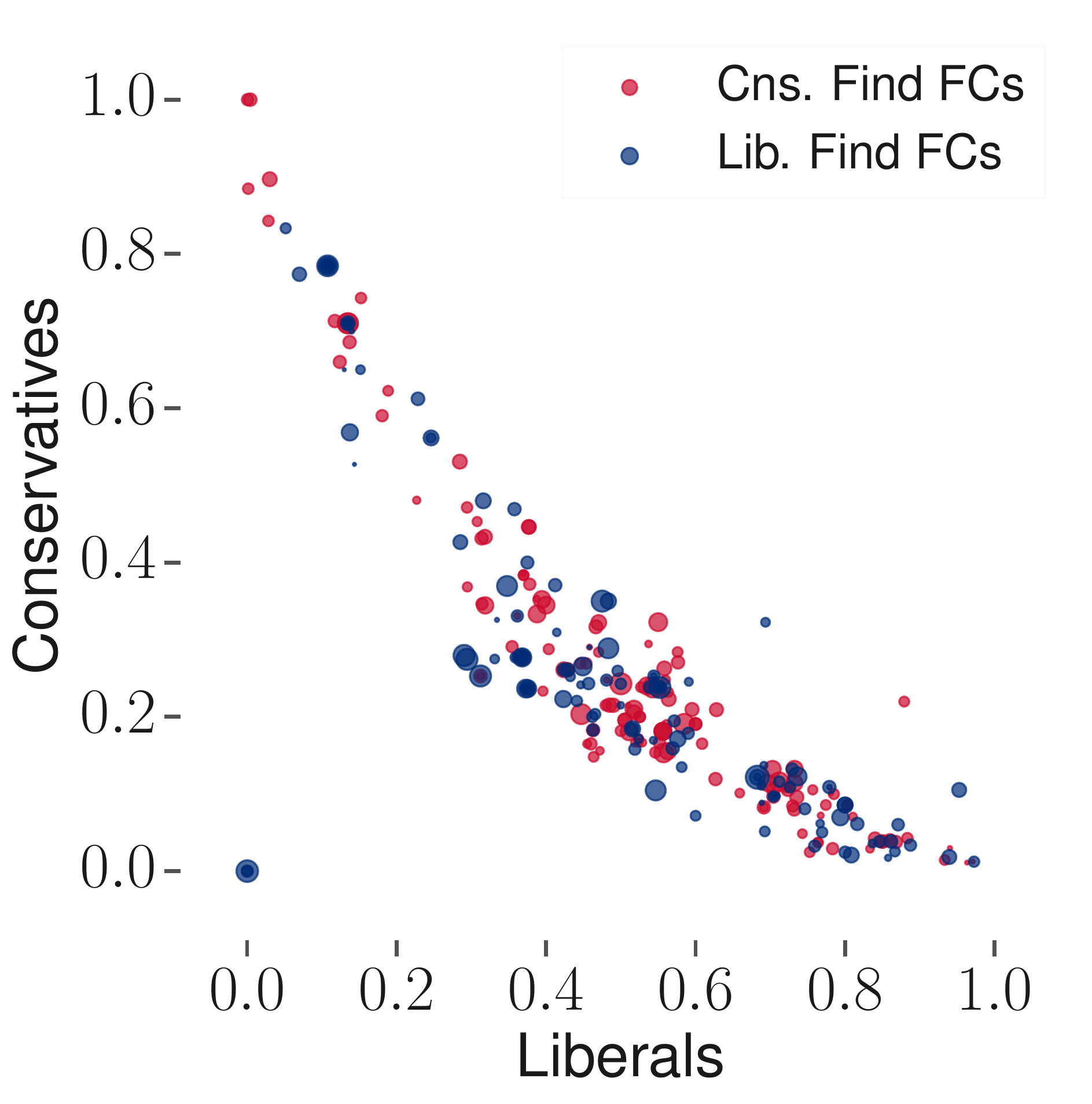}
    \caption{False Claims: Targeted}
    \end{subfigure}\hfill
    \begin{subfigure}[b]{0.5\columnwidth}
    \centering
  \includegraphics[width=\columnwidth]{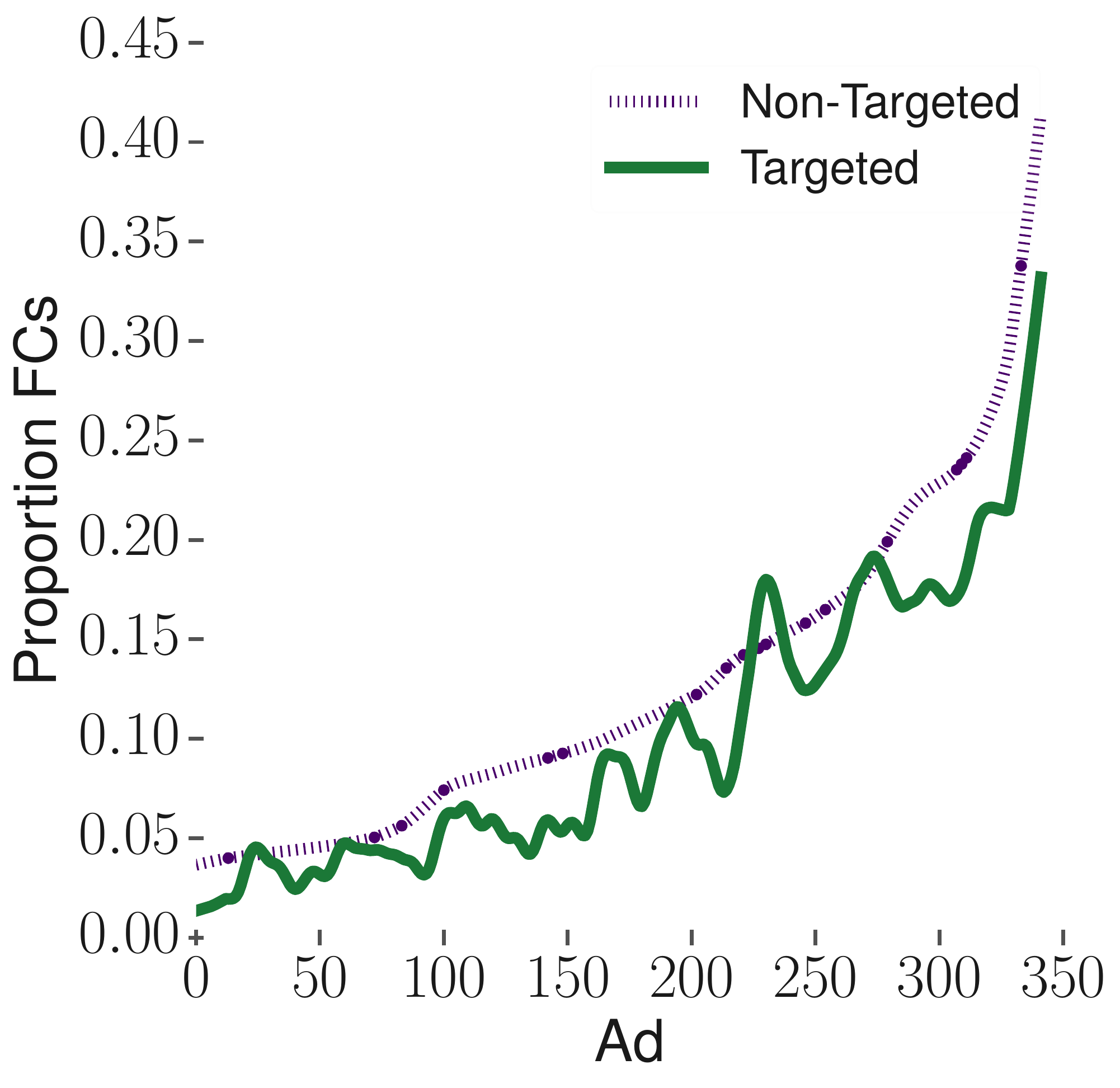}
    \caption{False Claims: Bias}
    \end{subfigure}\hfill
    \caption{Relationship between targeting and the responses by ideological groups. (a,c,e) show the proportion of population targeted and their tendency of response. Each circle represents an ad, and their size is proportionate with the between group disputability for that ad. (b,d,f) compares the mean responses of the targeted ads with their hypothetical non-targeted counterpart (i.e., overall responses), where each ad is represented on the $x$-axis}
    \label{fig:targetingPlots}
\end{figure}

To understand these nuances of targeted advertising, in this section, we focus on the relationship between the targeted population and the ideological divisiveness in reporting, approval, and false claim identifying behaviors for the ads. Table~\ref{table:correlationTargetingSummary} reports the correlation values between the targeted population and the tendency of the population to report, approve, and identify false claims. 

\textit{Reporting. } We observe a negative correlation in the case of reporting for both Liberals and Conservatives (also see Figure~\ref{fig:targetingPlots} (a)). This suggests that the targeted population has a lower tendency to report than the non-targeted one. This is also evident per Figure~\ref{fig:targetingPlots} (b), where we find that the reporting by the targeted population carries way lower likelihood than the reporting by the overall (or non-targeted) population.

\textit{Approval. } We observe a positive correlation in the case of approval for both Liberals and Conservatives (also see Figure~\ref{fig:targetingPlots} (c)). This suggests that the targeted population has a greater tendency to approve the ads as compared to the non-targeted population. This is also evident per Figure~\ref{fig:targetingPlots} (d), where we find that the approval score by the targeted population carries greater score for a majority of the ads compared to the overall (or non-targeted) population.

\textit{False claims. } For false claims, we do not find any significant correlation between the targeted population and divisiveness. However, per Figure~\ref{fig:targetingPlots} (e\&f) we do find that the targeted population has a lower tendency to identify false claims.

Taken together, we can assume that the ads were ``well-targeted'' in a way towards that population which was more likely to believe, and approve and subsequently less likely to report or identify false claims in them.

\subsection{Summary}

Our findings show that the IRA ads reached audiences that are very biased towards African-Americans and Liberals. More important, we show that ads were overall targeted towards a population that is more likely to believe, and approve and subsequently less likely to report or identify false claims in them.

\section{Concluding Discussion}

In this paper, we provide an in-depth quantitative and qualitative characterization of the Russia-linked ad campaigns on Facebook. Our findings suggest that the Facebook ads platform can be abused by a new form of attack, that is the use of targeted advertising to create social discord. These ads showed to be divisive, were 10 times more effective than a typical Facebook ad, were biased especially in terms of race and political leaning, and tended to be targeting more the users who are less likely to identify their inappropriateness. 
We also provide strong evidence that these  
advertisers have explored the Facebook suggestions tool to engineer the targeted populations. 


While this tool may be helpful in many ways, it needs to be carefully redesigned to avoid that a malicious advertiser reaches so easily groups of vulnerable people. 
For example, Facebook recently presented its intention to manually inspect ads before they are launched~\cite{fbBusiness2018},
aiming to guarantee that ads do not divide or discriminate people. Our work suggests that the priority of the candidates to be manually inspected can be based on their targeting formula. For instance, those ads that target extremely narrowed populations, on the basis of race, political leaning, and other sensitive topics have greater likelihood of being divisive. Additionally, the ads that experience severely high click-through rates could also be flagged to be quickly inspected. 

As a final contribution, we have deployed a system (available at \textit{\url{http://www.socially-divisive-ads.dcc.ufmg.br/}}) that displays the ads and their computed information such as the demographics of their targeting audiences.

\section{Acknowledgments}
F. Benvenuto and F. Ribeiro acknowledge grants from Capes, CNPq, and Fapemig. E. M. Redmiles acknowledges support from the U.S. National Science Foundation Graduate Research Fellowship Program under Grant No. DGE 1322106 and from a Facebook Fellowship. This research was partly supported by an European Research Council (ERC) Advanced Grant for the project ``Foundations for Fair Social Computing'', funded under the European Union's Horizon 2020 Framework Programme (grant agreement no. 789373). This research was partly supported by ANR through the grant ANR-17-CE23-0014. 

\bibliographystyle{ACM-Reference-Format}
\bibliography{Main}

\end{document}